\documentclass{aastex}
\usepackage{flushrt}

\slugcomment{To be submitted to the Astrophysical Journal}

\shortauthors{Limongi, Straniero, Chieffi}
\shorttitle{Massive Stars: Evolution and Nucleosynthesis}

\begin{document}

\title{MASSIVE STARS IN THE RANGE $\rm 13-25~M_\odot$:
EVOLUTION AND NUCLEOSYNTHESIS. II. THE SOLAR METALLICITY MODELS.}

\author{Marco Limongi\altaffilmark{1}, Oscar Straniero\altaffilmark{2} and Alessandro Chieffi\altaffilmark{1,3}}

\affil{1. Osservatorio Astronomico di Roma, Via Frascati 33, I-00040, Monteporzio Catone, Italy; 
marco@nemo.mporzio.astro.it} 

\affil{2. Osservatorio Astronomico di Collurania, I-64100, Teramo, Italy; straniero@astrte.te.astro.it}                   

\affil{3. Istituto di Astrofisica Spaziale (CNR), Via Fosso del Cavaliere, I-00133, Roma, Italy; 
achieffi@ias.rm.cnr.it} 

\begin{abstract}

We present the evolutionary properties of a set of massive stellar models
(namely 13, 15, 20 and 25 $\rm M_\odot$)
from the main sequence phase up to the onset of the iron core collapse. All these models have 
initial solar chemical composition, i.e. $Y=0.285$ and $Z=0.02$.
A 179 isotope network, extending from neutron up to $\rm ^{68}Zn$ and fully coupled to the evolutionary 
code has been adopted from the Carbon burning onward.
Our results are compared, whenever possible, to similar computations available in literature.

\end{abstract}



\section{Introduction}\label{intro}
This is the second paper in a series devoted to a detailed study of the evolutionary properties
of massive stars up to the onset of the final collapse. 
In the first paper of this series (\cite{CLS98} 1998 - hereinafter Paper I) we have
discussed in some detail the relevant literature on the subject
and the improvements progressively made in the computation of the advanced burning phases in these last two decades:
such a discussion will not be repeated here.
We discussed also the latest version of our evolutionary code FRANEC (version 4.2) together with
the evolutionary properties of a $\rm 25~M_\odot$ star, from the Pre Main Sequence up to the onset of the collapse. 
That evolution was computed 
by adopting a network which included 12 isotopes for the H burning, 25 isotopes for the He burning and
149 isotopes for all the more advanced burning phases.
Though such a network was well suited to present that first test model, an accurate tracing of the
chemical evolution of the matter requires a significantly larger nuclear network and hence
we now adopt in all our computations a much more extended network
(see section \ref{code}) with respect to the one adopted in Paper I.

In this paper we present the hydrostatic evolution of a first set of massive stars in the
mass range $\rm 13-25~M_\odot$ while in the next one of the series we will present
the evolution of metal poor ($Z=10^{-3}$ and $Z=0$) massive stars together with the explosive
yields. For the moment we 
did not extend the computations outside of this mass range because a)
stars less massive than $\rm 13~M_\odot$ are close enough to the limit of semidegenerate
C ignition that the computer time required to follow them becomes exceedingly large for the
computers we have at present and b)
stars more massive than $\rm 25~M_\odot$ begin to be significantly affected by mass loss,
phenomenon which we do not want to include yet.

Each subsection of section \ref{results} will be devoted 
to the analysis of a specific
burning (both central and in shell) and to its dependence on the initial mass.
Section \ref{disc} is devoted to a general discussion of the results together with a comparison
with similar computations available in literature.

A last summary and conclusions follows.

Since at the end of the writing of this paper the computation of all the other masses
of lower metallicity (i.e. $Z=0$ and $Z=10^{-3}$) were completed together with their explosive nucleosynthesis, since our
tables of explosive yields have been already distributed to many researchers, and since at present
they are also available at the web site \hfill\break
\centerline{\url{http://www.mporzio.astro.it/$^\sim$mandrake/orfeo.html}}
we decided to add our full set of the elemental explosive yields as appendix A to the present paper.

\section{Numerical methods and input physics}\label {code}
All the models have been computed by means of the evolutionary code FRANEC (ver. 4.2) whose
earliest and latest versions have been presented in \cite{CS89} (1989) and Paper I respectively.
The main properties of the code are the following.
The nuclear network includes 41 isotopes for the H burning, 88 isotopes for the He burning and 179
isotopes for the more advanced phases; a list of the isotopes adopted in these three regimes is shown in Table \ref{tabnet}.
\begin{table}
\dummytable\label{tabnet}
\end{table}
\placetable{tabnet}
The extension of the convective regions are fixed by means of the Schwarzschild criterion, and no
mechanical overshooting is allowed. On the contrary the semiconvection and the induced overshooting, due to the transformation
of {\rm He} in {\rm C} and {\rm O} during core helium burning, are properly taken into account
following \cite{CCPT85} (1985).
During the last phases of central {\rm He} burning the breathing pulses have been
inhibited as described in \cite{CCTCP89} (1989) and in \cite{CS89} (1989).
In the convective layers superadiabaticity was taken into account following
the derivation of \cite{CG68} (1968) of the mixing-length formalism of B\"ohm-Vitense.

The nuclear reaction rates are taken from the database kindly provided by Thielemann (private communication).
For the $^{12}{\rm C}(\alpha,\gamma)^{16}{\rm O}$
reaction we adopted the old value from \cite{CFHZ85} (1985).
The weak interaction rates as a function of the temperature and density
were taken from \cite{FFN80} (1980), \cite{FFN82} (1982) and \cite{FFN85} (1985). 
For the nuclei not included in this database, terrestrial half-lives 
have been adopted. The electron screenings are taken from \cite{GDGC73} (1973) and \cite{DGC73} (1973)
for the weak, intermediate and
intermediate-strong regime, and from \cite{ITI77} (1977) and \cite{ITID79} (1979) for the strong regime.

The equation of state (EOS) is the one described by \cite{Straniero88} (1988) and updated by \cite{SCL97} (1997). 

The radiative opacity coefficients are derived from \cite{Kurucz91} (1991),
\cite{IRW92} (1992) (OPAL) and from the Los Alamos Opacity Library (LAOL)
(\cite{Huebneretal77} 1977). The heavy element solar mixture by \cite{Grevesse91} (1991)
has been adopted.
The opacity coefficients due to the thermal conductivity are derived from \cite{Itohetal83} (1983).

The energy loss due to photo, pair and plasma neutrinos are properly taken into account following
\cite{Munakataetal85} (1985) (corrected as reported by \cite{Munakataetal86} 1986). 
Bremsstrahlung neutrinos are taken
into account following \cite{Dicusetal76} (1976) (corrected by \cite{Richardsonetal82} 1982) who extended the
results obtained by \cite{FestaandRudermann69} (1969) by the inclusion of the neutral current effects.
The energy loss due to the recombination processes are included following the prescriptions of
\cite{Beaudetetal67} (1967).

No mass loss has been included in the calculations.

\section{Evolutionary results}\label{results}
We followed the evolution of four stellar models of 13, 15, 20 and 25 $\rm M_\odot$,
having an initial solar chemical composition (i.e. $Z=0.02$ and $Y=0.285$), 
from the pre main sequence phase up to the onset of the iron core collapse. 
As usual the initial He abundance and the value
for the mixing length parameter ($\alpha=l/H_{\rm P}$) have been fixed by fitting the present 
properties of the Sun (see \cite{SCL97} 1997). The pre main sequence evolution has been
followed by assuming the star to settle on the Hayashi track directly with its final
main sequence mass.

The main evolutionary properties of all the computed models up to the precollapse stage (referred 
to, in the following, as the "final" model) are
summarized in Table \ref{tabmainprop}. 
\begin{table}
\dummytable\label{tabmainprop}
\end{table}
\placetable{tabmainprop}
The path followed by the star models in both the HR and the 
${\rm Log}T_{\rm c}$-${\rm Log}\rho_{\rm c}$ 
diagrams are shown in Figures \ref{fighr} and \ref{figtcroc}, while 
\placefigure{fighr}
\placefigure{figtcroc}
the evolution of the convective zones from the MS up to the central C ignition 
is shown in Figure \ref{figconv1};
\placefigure{figconv1}
Figure \ref{figragint} shows the temporal evolution of the radius of the internal mass layers
(in steps of $\rm 1~M_\odot$) from the central H exhaustion up to the presupernova stage.
\placefigure{figragint}

\subsection{H burning}\label{hburn}
The key properties of the four star models during central H burning are reported in Table \ref{tabmainprop} 
(section "H burning") from column (2) to column (5). The various quantities in column (1) refer to: 
the H burning lifetime in years ($\tau_{\rm H}$); the maximum extension of the H convective core in solar
masses ($\rm M_{\rm cc}$); the time duration of the H convective shell in years 
($\Delta t_{\rm H~conv~shell}$); the maximum extension of the H convective shell in solar masses
($\Delta M_{\rm H~conv~shell}$); the final location of the H burning shell (defined as the mass 
coordinate corresponding to the maximum of the nuclear energy generation rate provided by the
H burning reactions) in solar masses ($\rm M_{\rm H~shell}^{\rm final}$); the final location
of the He core (defined as the mass location corresponding to the H/He interface) in solar masses
($\rm M_{\rm He~core}^{\rm final}$); the final location of both the H burning shell and the He core
in solar radii (respectively $\rm R_{\rm H~shell}^{\rm final}$ and $\rm R_{\rm He~core}^{\rm final}$).

Central H burning proceeds, as it is well known, through the CNO cycle. The central temperature is high enough 
(between ${\rm Log}~T=7.5$ and ${\rm Log}~T=7.8$) for both the Ne-Na and the Mg-Al cycles
to become efficient:
$$\rm ^{20}Ne(p,\gamma)^{21}Na(\beta^{+})^{21}Ne(p,\gamma)^{22}Na(\beta^{+})^{22}Ne(p,\gamma)^{23}Na(p,\alpha)^{20}Ne$$
$$\rm ^{24}Mg(p,\gamma)^{25}Al(\beta^{+})^{25}Mg(p,\gamma)^{26}Al(\beta^{+})^{26}Mg(p,\gamma)^{27}Al(p,\alpha)^{24}Mg$$
As a consequence, at the end of H burning,
$\rm ^{21}Ne$ and $\rm ^{25}Mg$ are almost completely destroyed,
$\rm ^{22}Ne$ is reduced by almost an order of magnitude,
$\rm ^{23}Na$ and $\rm ^{26}Mg$ increase by a factor of $\sim 6$ and $\sim 2$ respectively,
$\rm ^{26}Al$ increases to a mass fraction of $\sim10^{-6}$ for the $\rm 13~M_\odot$ and $\sim10^{-7}$
for all the others, while $\rm ^{20}Ne$, $\rm ^{24}Mg$ and $\rm ^{27}Al$ remain almost unaltered. 
The central abundances of all these isotopes at the end of the H burning phase 
(specifically when the H mass fraction reduces to $10^{-3}$) are reported in Table \ref{tabcentral}
(section "H burning") for all the four models (columns 2 to 5).
\begin{table}
\dummytable\label{tabcentral}
\end{table}
\placetable{tabcentral}
All models form an H convective shell just before the central H exhaustion.
How it is well known (see e.g. \cite{LEB89} 1989), the efficiency of the mixing in this region 
determines if the star will ignite the He at the blue side or at the red side of the 
Hertzsprung gap: since we chose arbitrarily to adopt the Schwarzschild criterion, all our models
ignite He at the blue (quite obviously, the inclusion of mass loss in the computation
of these stars could in principle alter, even significantly, the path followed by the star in the HR diagram).
The further behavior of the surface properties of all the models
depends on the mass (see Figures \ref{figconv1}, \ref{figragint} and \ref{fighecteff}).
\placefigure{fighecteff}
In the two lower mass models the H convective shell vanishes rather soon and
hence the models early move towards the Hayashi track where most of the central He burning occurs.
The two more massive models show, on the contrary, a stable H convective shell 
which lasts for all the central He burning lifetime 
and hence they remain at the blue side of the Hertzsprung gap up to almost the end of the central He 
burning phase.

As each star approaches its Hayashi track, a convective envelope appears and penetrates
inward in mass. 
When the convective envelope reaches the mass location
marking the maximum extension attained by the H convective shell, it begins to dredge up
material modified by the H burning to the surface. Since both the 20 and $\rm 25~M_\odot$
models form a long lasting convective H burning shell which largely overlaps part of
the previous convective core, the matter dredged up to the
surface experiences two successive H burning episodes. 
On the contrary the matter dredged up to the surface by both the $\rm 13~M_\odot$ and the
$\rm 15~M_\odot$ was modified only by
the central H burning since in this case the H convective shell did not last enough
to induce a sizeable modification in the chemical composition.

The convective envelope reaches its maximum depth just before
the central neon ignition in all the four models; at this point the residual time before the explosion is
$\rm \simeq 145~yr$ for the $\rm 13~M_\odot$, 
$\rm \simeq 85~yr$ for the $\rm 15~M_\odot$, $\rm \simeq 77~yr$ for the $\rm 20~M_\odot$, and
$\rm \simeq 63~yr$ for the $\rm 25~M_\odot$. During this residual time the surface chemical 
composition remains practically frozen due to the more rapid
evolution of the core compared to that of the envelope.
Section "Convective Envelope" in Table \ref{tabmainprop} reports, for all the four models,
(column 2 to 5) the final locations of the inner boundary of the convective envelope in solar masses 
($\rm M_{\rm conv.~env}^{\rm final}$) and the final surface isotopic ratios 
$\rm ^{12}C/^{13}C$, $\rm ^{14}N/^{15}N$, $\rm ^{16}O/^{17}O$ and $\rm ^{16}O/^{18}O$.
The final surface abundances in mass fraction for all the nuclear species modified
by the H burning are reported in Table \ref{tabsurface} for all the four models (columns 3 to 6)
compared to the solar values (column 2). Note that in both the 20 and $\rm 25~M_\odot$ models
the convective H burning shell contributes to the final surface abundance of several species, i.e.
$\rm ^{16}O$, $\rm ^{17}O$, $\rm ^{23}Na$, $\rm ^{26}Mg$, $\rm ^{21}Ne$, $\rm ^{25}Mg$ and $\rm ^{26}Al$:
this last isotope in particular is almost completely produced by the convective H burning shell.

\begin{table} 
\dummytable\label{tabsurface}
\end{table}
\placetable{tabsurface}

The temporal behavior of both the mass and radius of the H burning shell (defined as the
mass corresponding to the maximum energy generation rate of the H burning) is shown in Figure \ref{figmrshell} as
a function of the central temperature up to the moment of the explosion. 
\placefigure{figmrshell}
The burning front advances in mass up to the end of the central
He burning phase and then it switches off completely when the formation of an efficient He burning shell
induces a robust expansion and hence the cooling of the H rich layers: note that the final radius
of the H shell scales inversely with the total mass, i.e., the larger the initial mass the smaller the final H shell radius
(see section \ref{disc}).

As a final comment on the behavior of the envelope, let us remark that the Kelvin Helmholtz timescale
for the envelope ($t_{KH}\simeq2\cdot10^{7}M_{\rm core}M_{\rm envelope}/RL_{\rm sup}$),
i.e. the timescale necessary to move (in the HR diagram) from the red to
the blue, and/or vice versa, is for all the four masses of the order of $\rm 15\pm5~yr$.
This means that (at least in the case of no mass loss) the position of all these stars in the HR diagram can
vary in principle up to $\sim 15$ yr before the explosion.

\subsection{He burning}\label{heburn}
The main properties of the central He burning for the four stellar models are reported 
in Table \ref{tabmainprop} (section "He burning"). The various quantities in column (1) are:
the time delay between the central H exhaustion and the central He ignition in years
($\Delta t(\rm H~exh.-He~ign.)$); the He burning lifetime in years ($\tau_{\rm He}$); 
the maximum extension of the He convective core in solar masses ($\rm M_{\rm cc}$); 
the central $\rm ^{12}C$ and $\rm ^{16}O$ mass fractions left by the central He burning;
the time duration in years and the maximum extension in solar masses of the first and the 
second He convective shells 
(respectively $\Delta t_{\rm 1~He~conv~shell}$, $\Delta M_{\rm 1~He~conv~shell}$, 
$\Delta t_{\rm 2~He~conv~shell}$ and $\Delta M_{\rm 2~He~conv~shell}$);
the final location of the He burning shell (defined as the mass 
coordinate corresponding to the maximum of the nuclear energy generation rate provided by the
He burning reactions) in solar masses ($\rm M_{\rm He~shell}^{\rm final}$); the final location
of the CO core (namely the He exhausted core) in solar masses
($\rm M_{\rm CO~core}^{\rm final}$); the final locations of both the He burning shell and the
CO core in solar radii (respectively $\rm R_{\rm He~shell}^{\rm final}$ and $\rm R_{\rm CO~core}^{\rm final}$).

This phase is characterized by the presence of a convective core which 
advances progressively in mass during all the central burning (see figure \ref{figconv1}) and by the conversion
firstly of He into $\rm ^{12}C$ and then of $\rm ^{12}C$ into $\rm ^{16}O$ via the $\rm ^{12}C(\alpha,\gamma)^{16}O$ 
reaction.
In addition to that, $\rm ^{14}N$ (which roughly equates the sum of the initial abundances of the nuclear species
previously involved in the CNO cycle) is completely converted into $\rm ^{22}Ne$ which,
in turn, is partially destroyed by the neutron producing reaction $\rm ^{22}Ne(\alpha,n)^{25}Mg$.

At variance with the He core mass, whose final value is determined by both the size of the convective
core in the H burning phase and by the advancing of the H shell in the following central
He burning phase, the final value of the CO core mass is determined only by the size of the convective
core at the end of the central He burning; the further evolutionary phases, in fact,
are fast enough that the main He convective shell which forms just outside the previous
border of the convective core does not have time enough to burn the available
fuel and hence to further advance in mass.
This means that the abundances of $\rm ^{12}C$ and $\rm ^{16}O$ within the whole CO core are
essentially flat and fixed by the (convective) central He burning phase and
not by a shell (radiative or convective) burning.
As it has been recognized a long time ago (see e.g. \cite{A72} 1972, \cite{TA85} 1985, \cite{WW86} 1986 and
\cite{WW93} 1993) the amount of carbon left 
by the central He burning phase strongly influences the further evolutionary phases
because it determines the amount of fuel available for both the central
and shell C burning (see next section).
It is important to stress here that the final amount of Carbon left by the central 
He burning depends on the efficiency of the $\rm ^{12}C(\alpha,\gamma)^{16}O$ 
reaction rate which, in turn, depends on both the amount of available fuel (mainly He)
and the nuclear cross section of the process:
in particular, either a larger nuclear cross section or a larger convective core
tend to lower the final carbon abundance.
This means that the same final abundance
of $\rm ^{12}C$ may be obtained by adopting a proper combination of different choices for both
the size of the convective core and the nuclear cross section. 
It is worthful to underline that both the extension of the convective core and the
$\rm ^{12}C(\alpha,\gamma)^{16}O$ cross section are still affected by large uncertainties.
It goes without saying that this degeneracy between convection
and nuclear cross section will be removed as soon as the ongoing 
experiments devoted to new measures of the nuclear cross section closer to the
Gamow peak will be completed.

A further difficulty is connected with the possible presence of the so called
Breathing Pulses. These instabilities occur during the last part
of the central He burning (typically when the central He mass fraction drops below 0.1)
and their effect is to induce an extra mixing which brings new fuel in the 
convective core. It exists a longstanding debate about the real existence of this
phenomenon (see \cite{CCTCP89} 1989 and \cite{CS89} 1989 for a comprehensive 
discussion on this subject)
which we do not want to address in this paper. We want simply to remark here that
the inclusion or the damping of these "Breathing Pulses" will significantly
alter the final carbon abundance.

For the moment, and for sake of clearness, let us clearly repeat that
the results shown in Table \ref{tabmainprop} have been obtained in the following scheme:
1) $\rm ^{12}C(\alpha,\gamma)^{16}O$ rate by \cite{CFHZ85} (1985),
2) Schwarzschild criterion (to check for convective instabilities),
3) inclusion of both the "induced" overshooting and semiconvection (see \cite{CCPT85} 1985),
4) inhibition of the Breathing Pulses if they occur.

As a final comment on the central He burning, let us note that while the final CO 
core scales directly with the mass, i.e. the larger the mass the larger the
final CO core, the $\rm ^{12}C/^{16}O$ ratio at the end of central He burning remains
almost flat in the mass range 15 to $\rm 25~M_\odot$
while it increases significantly in the $\rm 13~M_\odot$.

Once the He is exhausted at the center, the burning shifts outer in mass and locates at the He
discontinuity left by the convective core. The following evolution of the He burning is characterized
in all four cases by the formation of two successive, and overlapping, convective shells episodes: the first one
extends over a small mass range while the second one, which will last up to the collapse,
extends on a much larger zone. In particular the mass size of this second convective He shell is
mainly confined between the He discontinuity left by the convective core (i.e. the outer edge of the CO core)
and the He core mass.
Since the He core mass grows with the initial mass more rapidly than the CO core mass,
the mass size of the He convective shell increases with the initial mass of the star (see
Table \ref{tabmainprop}).
The final and typical chemical composition present in the He convective shell is obviously largely dominated
by He (more than 0.8 by mass fraction) and by carbon and oxygen. In the three more massive models the carbon abundance
in the convective shell is almost twice the oxygen one while in the smaller mass the oxygen abundance
is similar to the carbon one: this is due to the occurrence that in the $\rm 13~M_\odot$ the internal border of
the convective shell moves inward enough to bring  part of the oxygen left by
the previous central He burning within the convective shell. $\rm ^{14}N$ is fully
converted into $\rm ^{22}Ne$ while only a marginal fraction of it is converted in $\rm ^{25}Mg$.
All the main data concerning this phase are reported in Table \ref{tabmainprop}.

As a final remark let us note that, contrarily 
to the behavior of the H shell, the final radius of the He shell
scales directly with the initial mass (see Table \ref{tabmainprop}).

\subsection{The advanced evolutionary phases}\label{advphases}
The CO core left by the central He burning speeds up its contraction in order
to gain from the gravitational field the energy that the nuclear processes are not
able to provide any more. During this phase the core of these massive stars enters
a region of the $\rho-T$ plane in which neutrino emission (essentially pair 
production) becomes very efficient: such strong neutrino emission will continuously
increase all along the further evolutionary phases up to the final collapse.
The importance of the neutrino losses has been recognized long time ago (see, e.g., \cite{WAC72} 1972 and
\cite{WW86} 1986)
and in fact the advanced evolutionary phases have always been regarded as 
"neutrino dominated". Figure \ref{figvarielum} shows the temporal behavior (for each mass)
of the surface luminosity (energy lost by photons - dashed line), 
of the neutrino luminosity (energy lost by neutrinos - dotted line)
and of the nuclear luminosity (energy produced by all nuclear processes within the stars 
- solid line). 
\placefigure{figvarielum}
A comparison between the photon and neutrino luminosities clearly shows that
the main energy losses occur from the surface of the star up to the carbon 
ignition (see label C in Figure \ref{figvarielum})
and from the center in the more advanced burnings; since the nuclear
luminosity is regulated by the energy losses (it simply refurnishes the
star of the energy lost) it is clear that it closely follows the photon luminosity first
and the neutrino luminosity later. 
In the 20 and $\rm 25~M_\odot$ stellar models the nuclear
energy  during the advanced phases (after point C in Figure \ref{figvarielum})
 is not able to produce enough energy to fully sustain the star 
(the solid line runs below the dotted one) until 
the central Ne ignition: in these cases the contraction speeds up in order 
to gain the missing energy from the gravitational field. The various peaks which 
are visible in the nuclear luminosity mark the formation of convective episodes
(core or shell).

Let us now turn to a more specific analysis of the various
burning phases as a function of the initial mass of the star. 

\subsection{C burning}\label{cburn}
Carbon burning occurs few $10^{4}$ yr after the central He exhaustion.
This evolutionary phase totally depends
on the amount of carbon left by the central He burning since it
fixes the amount of 
available fuel. Moreover, since the neutrinos begin to be a major source
of energy loss from the center
of the star, also the formation of a convective core strictly depends on the initial carbon abundance:
in fact the minimum necessary (but not sufficient) requirement to form a
convective core is the existence of a positive flux, which implies that the
nuclear energy generation rate must exceed the neutrino losses;
since the nuclear energy generation
rate directly scales with the amount of fuel, it is clear that a larger initial carbon abundance favors the
formation of a convective core.

As we have already noted in section \ref{heburn}, three out of the four
models (15, 20 and 25) have a very similar central carbon abundance at the
end of central He burning; in spite of this similarity the $\rm 15~M_\odot$ develops 
a convective core which is not present in the other two more massive models (see Figure \ref{figconv2}).
\placefigure{figconv2}
The reason was already pointed out by \cite{A72} (1972) and, later on, by \cite{WW86} (1986)
and is that the neutrino losses
(pair production) scale inversely with the density while the nuclear energy production
scales directly with the density. Since the larger the CO core the lower the density
(at each given temperature), the balance between gains and losses becomes progressively
more negative as the CO mass increases.
As an example figure \ref{figcburnprop} shows a comparison between the $\rm 15~M_\odot$ (solid line) 
and the $\rm 25~M_\odot$ (dashed line). 
\placefigure{figcburnprop}
Panels a) and b) in Figure \ref{figcburnprop} show, respectively, the run of the central carbon abundance and of the 
density versus the central temperature,
while panels c) and d) show, respectively, the central run of the nuclear and 
the neutrinos energy generation rates. The vertical short dashed line marks
the temperature at which the 
convective core begins to develop in the $\rm 15~M_\odot$.
A comparison between panels c) and d) clearly demonstrates that the balance between
nuclear and neutrinos energy generation rates is negative for the more massive star
and positive for the smaller one: hence only the smaller of the two masses fulfill
the basic requirement for the formation of a convective core.

The main properties of the stellar models related to the C burning are reported in
Table \ref{tabmainprop} in section "C burning". The nuclear processes active in this phase
are roughly the same already
discussed in Paper I though the final most abundant nuclear species (see Table \ref{tabcentral})
differ from one model to the other because of the different conditions in which carbon burning occurs 
(i.e. temperature, density, convective core, etc.).
As the interplay between the local nuclear burning and the convective
mixing is concerned, we find that the typical mixing turnover time within the convective
core is of the order of $10^{-3}$ yr in both the less massive models, 
i.e. small enough to allow almost all the nuclear species to be homogeneously mixed.

Similarly to the previous evolutionary phases, 
once the carbon is exhausted in the center, a burning shell forms which
begins to move forward in mass.
Figure \ref{figcshellprop} shows 
the location of the C burning shell (solid line) superimposed to the convective zones as a function of time.
The behavior of this burning front and hence its final location
depends on the interplay among many factors, the main ones being
the amount of fuel which is encountered on the way out, the possible formation of convective shells and
the time available to the burning front to move further out. In our computations the C burning
shell advances in mass essentially until the central silicon ignition because at this point the time 
left before the explosion is so short (see Table \ref{tabmainprop}) that the 
C burning shell cannot significantly advance in mass any more. The mass size between the 
outer border of the last C convective shell and the He shell is negligible in the $\rm 13~M_\odot$ case while it
progressively increases up to $\rm 0.6~M_\odot$ for the $\rm 25~M_\odot$; this means that while in the smaller masses 
essentially no layer exposed to just the central He burning is preserved, for the larger masses a progressively
larger layer of material subject to only the central He burning (and hence the C/O ratio) 
is preserved up to the final explosion. 
\placefigure{figcshellprop}
Since the temporal behavior of the carbon convective shells may have important consequences for
the s-process nucleosynthesis, we report in Tables \ref{tabcshell1} and \ref{tabcshell2} some selected properties of the 
bottom of the carbon convective shells for two models, i.e. the $\rm 15~M_\odot$ and the $\rm 25~M_\odot$,
before and after each convective episode (four for the $\rm 15~M_\odot$ and two for the $\rm 25~M_\odot$).
\begin{table}
\dummytable\label{tabcshell1}
\end{table}
\placetable{tabcshell1}
\begin{table}
\dummytable\label{tabcshell2}
\end{table}
\placetable{tabcshell2}
During shell C burning,
the most efficient neutron producer is the $\rm ^{22}Ne(\alpha,n)^{25}Mg$ reaction which is almost always
of the order of $\sim10\%$ of the maximum integrated flux. By the way let us remind that the integrated flux 
of a nuclear process is defined as the integral of its reaction rate over a given time interval. 
Note that in Tables \ref{tabcshell1} and \ref{tabcshell2} the reaction having the
maximum integrated flux (i.e., the most efficient one) changes as a function of the mass.
The most efficient neutron poisons are basically the
$\rm ^{25}Mg(n,\gamma)^{26}Mg$,
$\rm ^{24}Mg(n,\gamma)^{25}Mg$,
$\rm ^{23}Na(n,\gamma)^{24}Na$ and
$\rm ^{27}Al(n,\gamma)^{28}Al$ reactions, whose relative and absolute efficiencies vary from
one model to the other and during the various convective episodes.
Note that the process $\rm ^{20}Ne(n,\gamma)^{21}Ne$, though it is very efficient, cannot
be considered as a neutron poison, but at most as a modulator of the neutron flux,
since it is immediately followed by the $\rm ^{21}Ne(\alpha,n)^{24}Mg$ which reemits
the previously absorbed neutron.
Within each convective shell $\rm ^{12}C$ and $\rm ^{22}Ne$ are partially destroyed while the
abundances of  $\rm ^{20}Ne$, $\rm ^{23}Na$, $\rm ^{24}Mg$, $\rm ^{25}Mg$ and $\rm ^{27}Al$ are
slightly increased by an amount which varies from one model to the other and depending on the
convective episode. The temperature and density at the bottom of the convective shell
vary between $\rm 7.8\cdot10^{8}~K$ and $\rm 3.3\cdot10^{5}~gm/cm^{3}$
(the first convective episode of the $\rm 15~M_\odot$ model)
and $\rm 1.4\cdot10^{9}~K$ and $\rm 1.14\cdot10^{5}~gm/cm^{3}$
(the last convective shell of the $\rm 25~M_\odot$ model). Note that the temperature increases while
the density decreases from a convective episode to the following one, the temperature being larger
and the density being lower with increasing the mass of the star.
For each model the maximum neutron density (number of free electron per unit volume) increases from a convective
shell to the following one; during each convective episode, for each mass coordinate, it fastly increases 
to its maximum value and then it lowers following an almost exponential low.

The final location in mass and radius of the carbon burning shell, in all the computed models, 
is reported in Table \ref{tabmainprop}.

\subsection{Ne burning}\label{neburn}
As the central temperature reaches $\rm \simeq1.3\cdot10^{9}~K$ Ne burning takes place. The
time which elapses between the end of the central carbon burning and the onset of the Ne burning ranges
between 1500 yr (for the $\rm 13~M_\odot$) and  92 yr (for the $\rm 25~M_\odot$). 
At this stage the carbon burning shell is located,
for the four models, at $\rm 1.40~M_\odot$ $(\rm 13~M_\odot)$, $\rm 1.48~M_\odot$ $(\rm 15~M_\odot)$,
$\rm 1.46~M_\odot$ $(\rm 20~M_\odot)$ and $\rm 1.51~M_\odot$ $(\rm 25~M_\odot)$ and the number of
convective C shell episodes already experienced up to this moment are:
three ($\rm 13~M_\odot$ and $\rm 15~M_\odot$), two $(\rm 20~M_\odot)$ and one $(\rm 25~M_\odot)$
(see Figures \ref{figconv2} and \ref{figcshellprop}).
Neon burning occurs in a convective core in all the four models since the amount of available fuel is
always sufficient for the nuclear energy production to overcome the neutrino losses.
The neon abundance almost equates the preexisting carbon abundance (by number) since it is 
its main product and hence it directly depends on the choices 
made for the He burning (see section \ref{heburn}).

The main properties of the central neon burning for all the computed models are reported in Table \ref{tabmainprop},
while the main nuclear species left at the end of the central neon burning (i.e. the ones whose mass fraction is
greater than 0.01) are reported in Table \ref{tabcentral}.
The dominant processes involved in this burning have also been discussed in Paper I and will not be
repeated here.

As the central neon mass fraction drops below $\sim10^{-4}$ the neon burning shifts in a shell.
The following evolution of the burning front strictly follows the fate of the inner core
in the sense that it almost stops its advancing during the main burnings (central
and shell oxygen burnings and central silicon burning) while it significantly advances in mass 
between two consecutive burnings. The reason is that the "ensemble" of all the more advanced burnings lasts so short
that the neon shell does not have time to advance: only between two successive burnings the cores
experience a strong contraction and heating which induces a sudden burning (and hence
advancing) of the Ne shell.
The situation is discussed in detail in Paper I and occurs in the same way
in all the models presented here.

A neon convective shell forms
only in the $\rm 15~M_\odot$ just after the central oxygen exhaustion; it is located at a mass 
coordinate of $\rm 0.93~M_\odot$ and
reaches its maximum extension between 0.93 and $\rm 1.24~M_\odot$ in $\rm \simeq0.05~yr$
when the local $\rm ^{20}Ne$ mass fraction is reduced to $\sim10^{-4}$ and then it recedes in mass.

The final location in mass and radius of the neon burning shell is reported in Table \ref{tabmainprop}.

\subsection{O burning}\label{oburn}
Central oxygen burning starts just after the central neon exhaustion. It always occurs
in a convective core since the nuclear energy production
is always large enough to overcome the neutrino losses. 
At variance with the previous central burnings, the size of the convective core 
does not depend significantly on the mass but it is always
of the order of $\rm 1~M_\odot$ (see Table \ref{tabmainprop}). The central O burning lifetime 
ranges between 8 yr for the $\rm 13~M_\odot$ and 0.33 yr for the $\rm 25~M_\odot$. 
The timescale of this burning
is short enough that all the various burning shells (Ne, C, He and H) remain practically
freezed out during this phase.
Since the more advanced evolutionary phases are even faster, the various shells (apart from
the Ne one) will remain freezed out up to the moment of the final collapse.
The Ne shell, as already addressed in the previous section, is the only exception because 
it is located in a zone which undergoes a strong
contraction and heating after the central oxygen burning and hence it is the only one which may 
significantly advance in mass anyway. 

The most efficient nuclear processes are fairly the same already discussed in Paper I 
even if the final nucleosynthesis (Table \ref{tabcentral}) depends appreciably on the initial mass. 
In particular it happens that while $\rm ^{28}Si$, $\rm ^{32}S$ and $\rm ^{38}Ar$ 
are the most abundant nuclei in the two more massive models, in the two less massive ones
the two most abundant nuclei, other than $\rm ^{28}Si$, 
become $\rm ^{30}Si$ and $\rm ^{34}S$, i.e. more neutron rich nuclei. This is due to the well known
occurrence that during central oxygen burning many weak processes become efficient 
(in particular the processes $\rm ^{31}S(\beta^{+})^{31}P$, $\rm ^{33}S(e^{-},\nu)^{33}P$, 
$\rm ^{30}P(e^{-},\nu)^{30}Si$ and $\rm ^{37}Ar(e^{-},\nu)^{37}Cl$) and that 
the efficiency of these processes scales inversely with the mass.
Table \ref{tabweakoxy} shows the integrated flux 
of the main weak processes during the central O burning for two selected masses, namely
the 15 and $\rm 25~M_\odot$. In Table \ref{tabweakoxy}, the electron fraction $Y_{\rm e}$
refers to the end of oxygen burning.
\begin{table}
\dummytable\label{tabweakoxy}
\end{table}
\placetable{tabweakoxy}

Once oxygen is exhausted at the center, the burning moves outward and a first (main) O convective shell 
develops (in all the models), at a mass location corresponding roughly to the previous maximum 
extension of the O convective core (see Figure \ref{figconv2}).
The time duration and maximum extension reached by this first convective episode is reported in 
Table \ref{tabmainprop} for the four masses.
Once this first O convective shell vanishes, a second one forms at a mass 
coordinate more internal than the previous one; the reason has already
been discussed in detail in Paper I (see section 3.5 and Figure 13 of Paper I) for the 
$\rm 25~M_\odot$ and holds also for the 13, 15 and $\rm 20~M_\odot$. The last O convective 
shell forms almost simultaneously to the first silicon convective shell, 
the only exception being the $\rm 13~M_\odot$ model.
The duration of this last oxygen convective
episode is rather short, i.e. of the order of $10^{-4}$ yr (see Table \ref{tabmainprop}) but it is 
rather important since it extends up to almost the base of the C burning shell, with
the consequence that the Ne shell is pushed outward and squeezed towards the C one.
The final location both in mass and radius of the oxygen burning shell is reported in Table \ref{tabmainprop}.

\subsection{Si burning}\label{siburn}
Silicon burning takes place when the central temperature reaches about $2.3\cdot10^{9}~{\rm K}$.
The time elapsed since the central oxygen exhaustion ranges between
160 days for the $\rm 13~M_\odot$ and 11 days for the $\rm 25~M_\odot$. The degree of neutronization at the
beginning of the Si burning (see Table \ref{tabcentral}) scales inversely with the initial mass of the stars (because
both the density and the evolutionary timescales scale inversely with the masses) and constitutes
an important "boundary condition" since the Si burning
is largely affected by such an initial neutronization (it controls at a large extent the nuclear
species which will be favored in the further evolutionary phases).

Quite in general, the central Si burning phase may be divided in two phases: the first one
in which Si burns radiatively and a second one in which it occurs in a convective environment.

In the $\rm 13~M_\odot$ case the first "radiative" phase is characterized by the total 
destruction of $\rm ^{28}Si$ and by a build up of $\rm ^{30}Si$ up to a maximum mass fraction 
of more than 0.8. At this point a convective core appears and the abundance of $\rm ^{28}Si$
increases again as a consequence of the advancing of the convective core in a region in 
which $\rm ^{28}Si$ is abundant (it is still the one left by the O convective core). 
However the $\rm ^{28}Si$
abundance never exceeds the $\rm ^{30}Si$ one, which remains the most abundant silicon isotope
up to the end of this burning phase, because of the high degree of neutronization existing within the 
convective core. The maximum mass size of the convective core is $\rm \simeq 1~M_\odot$ and the matter emerging from
this burning is formed by more than 70\% (by mass)
of $\rm ^{52}Cr$ and $\rm ^{56}Fe$, the first one being largely the most abundant of the two.
For sake of completeness we report in Table \ref{tabcentral} the abundances of the most abundant nuclear species
during the central Si burning at three selected points: the end of the radiative phase, the
moment at which $\rm ^{28}Si$ reaches its maximum abundance during the convective phase and the 
end of the central Si burning phase. The electron fraction is also reported.

As we move upward  in mass, i.e. to the $\rm 15~M_\odot$, the development of the Si burning changes 
as a consequence of the reduction in the average neutronization. 
During the first radiative burning, $\rm ^{28}Si$ is largely destroyed (but no more completely)
and $\rm ^{30}Si$ is builded up to a maximum mass fraction of $\simeq0.675$. Once the convective core
forms and begins to advance in mass, it enters a region in which the $\rm ^{28}Si$ is abundant and
the neutron excess is low: the result is now that the $\rm ^{28}Si$ grows up to a value of $\simeq0.347$,
becoming the most abundant Si isotope. During all the convective phase
$\rm ^{30}Si$ remains largely underabundant with respect to $\rm ^{28}Si$. 
Once again at the end of central Si burning, most of the matter 
($\simeq83\%$ by mass) is concentrated in $\rm ^{52}Cr$ and $\rm ^{56}Fe$ but this time the two
nuclear species have more or less the same abundance by mass. The detailed (main) abundances 
at selected points along the Si burning are shown in Table \ref{tabcentral}. The maximum size reached by
the convective core is, also in this case, $\simeq 1~M_\odot$.

The trend which comes out by a comparison between the evolution of the 13 and the $\rm 15~M_\odot$ 
is confirmed and reinforced by an analysis of the Si burning in the two more massive stellar
models, i.e. the 20 and the $\rm 25~M_\odot$. In fact, as the stellar mass increases, the amount of $\rm ^{28}Si$
burned during the first radiative phase reduces progressively 
so as the maximum abundance reached by the $\rm ^{30}Si$.
In particular $\rm ^{30}Si$ grows up to 0.25 in the $\rm 20~M_\odot$ and only up to 0.2 in the $\rm 25~M_\odot$, 
while $\rm ^{28}Si$  is only mildly depleted in this phase. As soon as the convective core forms, 
$\rm ^{30}Si$ almost disappears as a consequence of the low degree of neutronization existing within the whole core 
of both the masses. The final composition is dominated once again
(more than $\simeq85\%$ by mass) by $\rm ^{52}Cr$ and $\rm ^{56}Fe$, the iron isotope
being now the most abundant of the two (see Table \ref{tabcentral}). Note that also these two stellar models
form a convective core of the order of $\simeq 1~M_\odot$.

The main weak interactions which drive the progressive neutronization of the matter from the central
Si ignition onward are reported in Table \ref{tabweaksi}
for two representative cases, i.e. the 15 and the $\rm 25~M_\odot$ models.
In particular we report in the table, for each process, the integrated effective flux and the $\varphi$ parameter.
The integrated effective flux is defined as:
\begin{equation}\label{fi}
\int{ \vert r_{ik}-r_{jl} \vert} dt 
\end{equation}
where $r_{ik}$ and $r_{jl}$ are the forward and reverse rate of the process $i(k,j)l$; such a quantity
is an estimate of the global efficiency of a certain process over a given time interval.
The $\varphi$ parameter is defined (see also Paper I) as:
\begin{equation}
\varphi(ij)={\vert r_{ik}-r_{jl} \vert \over {\rm max}(r_{ik},r_{jl})}
\end{equation}
and gives an estimate of the balance between the forward and reverse rate of the process $i(k,j)l$.
\begin{table}
\dummytable\label{tabweaksi}
\end{table}
\placetable{tabweaksi}
For sake of clearness Table \ref{tabweaksi} is divided into three panels: the first one refers to the
central convective Si burning, the second one concerns the phase extending from
the central Si exhaustion up to the onset of the Nuclear Statistical Equilibrium (NSE) and
the last one refers to the final phase up to the last computed model.
The last row at the end of each panel shows the process having the maximum 
integrated effective flux.
Let us note that the most efficient weak processes during these phases are always electron
captures on heavy nuclei and that these processes are far away from the equilibrium before the NSE
is reached. Only during the last evolutionary stages some of these weak reactions begin to be counterbalanced
by their reverse process like, e.g., the $\rm ^{52}Cr(e^{-},\nu)^{52}V$ in the $\rm 15~M_\odot$ or 
the $\rm ^{59}Co(e^{-},\nu)^{59}Fe$ in the $\rm 25~M_\odot$. 

Once Si is exhausted in the center the burning shifts in a shell.
In spite of the very short time available for these models before the beginning of the final collapse
(and, possibly, a successful explosion),
almost 6 hours for the $\rm 13~M_\odot$ and only 0.5 hours for the $\rm 25~M_\odot$, 
the Si burning shell has time to move significantly outward in mass and even to produce (in all the four masses) 
a series of subsequent convective episodes.
The first one (or two, depending on the initial mass) occurs always
within the maximum extension previously reached by the convective core and is due to the
presence of the small amount of Si left by the receding convective core. The following one(s) extends,
on the contrary, well outside of the border of the convective core.
The maximum extension in mass reached by the outer border of this convective Si shell is a very crucial quantity
since it marks the discontinuity which divides the highly neutronized matter (the internal one)
from the almost untouched one (the external one).
This discontinuity, usually referred to as the
"Fe" core mass, plays a pivotal role in the following evolution of the star, i.e. during the 
explosive phase. In fact it marks the limiting mass within which the shock wave looses
a great deal of its energy ($\rm \sim10^{51}~erg/0.1~M_{\odot, \rm Fe}$, see \cite{HM81} 1981
and references therein for a more detailed discussion)
through the photodisintegration of heavy nuclei. This means that the smaller is the "Fe" core mass the largest  
is the probability of obtaining a successful explosion. Secondly, since even a mild neutronization ($Y_{\rm e}\simeq0.49$)
leads always to an NSE distribution which has a negligible abundance of $\rm ^{56}Ni$, the region outside
the "Fe" core is the only one in which a large amount of $\rm ^{56}Ni$ may be produced during the explosion.
We find that in all the three more massive models the maximum extension in mass of the convective shell 
is rather similar (it varies between $\rm 1.43~M_\odot$ and $\rm 1.55~M_\odot$) while in the 
$\rm 13~M_\odot$ model the convective
shell only marginally extends above the previous convective core, so that the final "Fe" core mass is
$1.29~M_\odot$.

The nucleosynthesis occurring in the shell Si burning phase is basically
the same already discussed in Paper I for the $\rm 25~M_\odot$, i.e., the Si burning shell leaves a
chemical composition mainly enriched in $\rm ^{54}Fe$ and $\rm ^{56}Fe$ in proportions which differ
from one model to the other; the difference between the central and the
shell Si burning is mainly due to a difference in both the entropy and the neutron excess
at which the nuclear burning occurs (see Paper I for more details). 

The final location in mass and radius of the silicon burning shell is reported in Table \ref{tabmainprop}.

\section{Discussion}\label{disc}
The interplay between the various central and shell nuclear burnings and the temporal behavior
of all the convective zones leads to final presupernova structures whose main properties 
are shown in Figures \ref{figmassrag}, \ref{figmassdens}, \ref{figmassye}, 
\ref{figmassentr}, \ref{figvariepress} and \ref{figmcores} and
in the last section of Table \ref{tabmainprop}.

\placefigure{figmassrag}
\placefigure{figmassdens}
\placefigure{figmassye}
\placefigure{figmassentr}
\placefigure{figvariepress}
\placefigure{figmcores}

Figures \ref{figmassrag} and \ref{figmassdens} show, respectively, the final radius-mass and density-mass relations obtained
for the four stellar models; how it has been already shown by \cite{A96} (1996), these relations tend to progressively 
approach each other as one moves towards the center. This simply means that the more advanced the evolutionary
phase the lesser is the influence of the initial mass. 
This occurrence is evident also in Figure \ref{figmcores} where the various final core masses
(or, equivalently, the final location of the various burning fronts) are plotted as a function
of the initial mass: the more internal core masses show clearly a shallower dependence on the initial
mass than the more external ones.
For example, while the final location of the H shell (i.e. the He core mass)  
in the $\rm 25~M_\odot$ is about 2.4 times larger than in the $\rm 13~M_\odot$,
the mass location of the Si shell (i.e. the most internal one) in the 
$\rm 25~M_\odot$ is only 1.4 times larger than that of the $\rm 13~M_\odot$.

Figure \ref{figmassye} shows the final electron fraction ($Y_{\rm e}$) profile within the inner $\rm 2~M_\odot$ 
for all the computed models. A main jump in $Y_{\rm e}$ is present between 1.3 and $\rm 1.55~M_\odot$ 
depending on the initial
mass of the star; it is determined by the outer border of the convective Si burning shell since the region
affected, even partially, by the Si burning experiences a rather strong neutronization. The mass within this
main jump in $Y_{\rm e}$, i.e. the iron core, is presumed to remain locked within the remnant
on the basis of nucleosynthetic restrictions (see \cite{WW78} 1978 and \cite{WW95} 1995 for more details).
The region outside the iron core mass, on the contrary, will be in large
part ejected in the interstellar medium by the explosion. Actually the exact location marking the region 
which will be successfully ejected is at present largely unknown and constitutes one of the main uncertainties 
in the computation of the final explosive yields (see, e.g., \cite{TNH96} 1996). 
The $Y_{\rm e}$ profile which is present outside
the iron core mass is due to the superposition of two successive episodes of neutronization. The first one
occurs during the He burning and is due to the transformation of $\rm ^{14}N$, left by the H burning, into  
$\rm ^{22}Ne$: this
first episode leaves a $Y_{\rm e}$ profile almost flat within all the region involved by the He burning and it is independent on
the initial mass of the star (in fact the amount of $\rm ^{22}Ne$ produced is always equal to the total
amount of $\rm ^{14}N$ available which, in turn, equates the initial global abundance of the isotopes involved in the CNO
cycle which obviously depends only on the initial metallicity). The second episode occurs during the convective
O burning shell since both the C and Ne burnings do not produce any appreciable amount of neutronized
matter. At
variance with the first episode this time the degree of neutronization is a strong function of the 
initial mass and largely decreases as the initial mass increases.
In both our two more massive stars the O shell burning does not produce an appreciable further neutronization
and hence the $Y_{\rm e}$ profile remains practically flat outside the iron core mass. 
In the two less massive stars, on the contrary, the convective O burning shell
induces an appreciable amount of neutronization; hence the final $Y_{\rm e}$ profile outside the iron core will
depend on the mass extension of both the O and the Si convective shells. In particular
in the $\rm 13~M_\odot$ case we find that the last Si convective episode is not able to extend up to the outer border
of the previously existing O convective shell and hence part of the partially neutronized matter
produced by the O shell remains outside the iron core. 
In the $\rm 15~M_\odot$ case
the last Si convective shell extends above the previous O convective shell and hence in this case
the partially neutronized matter produced by the O shell is fully engulfed by the advancing Si shell.
As a consequence, in this case, the $Y_{\rm e}$ profile outside the iron core
is almost flat.

For sake of completeness Figures \ref{figmassentr} and \ref{figvariepress} show, respectively, the 
specific entropy per baryon ($S/N_{\rm A}k$) and the 
various contributions to the total pressure (i.e. radiation, ions, electrons and Coulomb interactions) 
for the four presupernova models. Note that the specific entropy is computed only for $\rm T>10^{6}~K$ 
where the matter is assumed to be completely ionized. Figure \ref{figvariepress} clearly shows that within the
iron core the star is supported mainly by the degenerate electrons while outside the iron core
the radiation contributes, or even dominates, in sustaining the stellar model: the contribution of the
Coulomb interactions to the total pressure is in all four cases largely negligible.

Turning now to the chemical structure of the four presupernova models,
Figure \ref{figmajors} shows the profiles of the main isotopes as a function of the
internal mass; 
\placefigure{figmajors}
it can be easily recognized the onion structure left by the various burnings and it is also possible
to determine the mass layers, summarized in Table \ref{tabkeyzones}, which experienced the nuclear burnings up to 
a given one. 
\begin{table}
\dummytable\label{tabkeyzones}
\end{table}
\placetable{tabkeyzones}
To be more specific, the label to the left of Table \ref{tabkeyzones} reports up to which burning (included) the
matter between the two bracketing masses on the right has been exposed. For example the mass between
$\rm 1.754~M_\odot$ and 
$\rm 1.643~M_\odot$ in the $\rm 15~M_\odot$ has been processed up to the Ne burning included. Note that the zone affected
by the central H burning extends up to the surface because of the effect of the dredge up.
Tables \ref{tabyield13}-\ref{tabyield25} schematically show the percentage of the abundance of each isotope present within 
each of the mass zones reported in Table \ref{tabkeyzones}. Note that the sum of the percentages does not always 
equal 100 because the decimal part of each fraction has been rounded off in the printout format.
The last column in each of these four tables shows the final preexplosive yield of each
isotope computed up to the border of the iron core. 
\begin{table}
\dummytable\label{tabyield13}
\end{table}
\placetable{tabyield13}
\begin{table}
\dummytable\label{tabyield15}
\end{table}
\placetable{tabyield15}
\begin{table}
\dummytable\label{tabyield20}
\end{table}
\placetable{tabyield20}
\begin{table}
\dummytable\label{tabyield25}
\end{table}
\placetable{tabyield25}

A comparison between the present evolutions and similar computations
available in literature is a very difficult task since most of the
work devoted to the evolution of massive stars contains very few details
on the presupernova models. The comparison between the results obtained
by different authors is almost always limited to the final, explosive,
yields. However, by looking in the literature of the last decade we were able to collect
some data on the presupernova models of massive stars; they refer,
essentially, to: \cite{NH88} (1988) (hereinafter NH88), \cite{WW95} (1995) (hereinafter WW95), 
and \cite{APB96} (1996) (hereinafter APB96).

Figure \ref{figcompmcore} shows the comparison of the final location in mass of the various
nuclear burning shells for the 13, 15, 20 and $\rm 25~M_\odot$, with WW95 (filled triangles),
NH88 (filled squares) and APB96 (asterisk). In the following we will simply underline the similarities
and the differences among the different sets of models. A real explanation of the causes of the
differences existing among the various sets of models would be a too tough job
either because of the lack of published data 
and because of the intrinsic difficulty in performing such an analysis.
\placefigure{figcompmcore}
The upper left panel in this figure collects the final He core masses obtained by the quoted papers. This quantity is of
paramount importance since, once the H is exhausted in the center, the evolution of a stellar model 
depends essentially on the mass size of the He core and not any more on the total mass of the star.
The comparison
between our results and the NH88 ones is not meaningful because they start their computations from pure He cores;
however, the fact that both of us have similar He cores will allow
a meaningful comparison of the more advanced evolutionary 
phases. The $\rm 20~M_\odot$ stellar model computed by APB96 closely matches our
corresponding He core mass, and this agrees with
the fact that both of us adopt the Schwarzschild criterion to define the border of the convective zones. The comparison
with WW95 is somewhat controversial in the sense that their $\rm 15~M_\odot$
is in close agreement with the present one while their $\rm 25~M_\odot$ forms a significantly larger core 
(by $\rm \simeq1~M_\odot$) respect to our.
Since WW95 adopt a mild amount of overshooting ($\simeq0.25~H_{\rm P}$) we would expect their final He core masses to be systematically
larger than ours. Though one could imagine the overshooting to be more effective in the 
$\rm 25~M_\odot$ rather than in the $\rm 15~M_\odot$, some test computations we have performed seem to exclude such a possibility.

The upper right panel in figure \ref{figcompmcore} shows the comparison among the final CO core masses.
The NH88 CO core masses for the 13, 15 and $\rm 20~M_\odot$ models closely match our findings
while they obtain a significantly larger CO core for the $\rm 25~M_\odot$ star in spite of the fact that
both have a similar He core and none of us includes the mechanical overshooting.
APB96 obtain a significantly larger CO core (for their $\rm 20~M_\odot$); this occurrence can be easily explained
since they clearly state that they find, and do not dump, some breathing pulses (\cite{CCPT85} 1985 and
\cite{CCTCP89} 1989) whose main effect, as it is well known, is to increase both the mass of the
CO core and the He burning lifetime. WW95 obtain larger CO cores as a result of the larger He core masses and of the 
larger He convective cores due to the mechanical overshooting adopted during the central He burning.

The final mass location of the C burning shell coincides with the base of the last carbon
convective shell where the nuclear energy production due to carbon burning reaches
its maximum. It is worth noting, however, that the main changing from a C rich layer (i.e. the CO core) to
a Ne rich one (i.e. the ONe core) occurs at the outer border of the last C convective shell
(see Figure \ref{figmajors}).
By the way let us note that, in general, while the final structure of the star depends on the location in mass
of the active burning shells (as discussed above), the final stellar yields are sensitive to the  size of the 
various core masses (see below). From the
middle left panel of Figure \ref{figcompmcore}, in which the final location of the ONe core (rather than
the location of the C burning shell) is plotted, it is clear that WW95, NH88 and APB96 are in a rather good 
agreement among them.
Our lower mass models (namely the 13 and the $\rm 15~M_\odot$) are in agreement with both WW95 and NH88,
while our higher masses show a significantly smaller ONe cores. This is due to the fact that
the our two more massive stellar models form systematically a smaller C convective shell (Table \ref{tabcconvshell}).
\begin{table}
\dummytable\label{tabcconvshell}
\end{table}
\placetable{tabcconvshell}
The behavior of the carbon convective shell directly influences the temporal evolution of the Ne burning
shell since it acts as a barrier for the advancing Ne shell itself. The large difference between the $\rm 25~M_\odot$ star
by WW95 and our corresponding model seems to indicate that in the WW95 model the carbon convective shell vanishes at
a certain point of the evolution and allows the neon burning shell to advance in mass. On the contrary
in our $\rm 25~M_\odot$ the carbon convective shell is active up to the end of the evolution keeping
the Ne shell more internal. NH88 give no data on the location in mass of the Ne shell, however by comparing the 
final location of the O burning shell it is possible to guess that NH88 obtain a temporal evolution of the carbon 
convective shell similar to the one we found. APB96 stop the evolution of their $\rm 20~M_\odot$ to the end of
central Ne burning and hence do not have data for the final location of both the Ne and O burning shells.
The comparison among the various final mass locations of the O shell shows that while the NH88 models fairly
agrees with the present ones, the WW95 models end up with an O shell systematically more external than in our models.

A last important quantity we can compare is the final size of the iron core, which is directly connected
to the extension in mass of the last silicon convective shell (see Paper I for more details).
A comparison with NH88 shows that we obtain iron cores for the 13, 15 and $\rm 20~M_\odot$,
systematically more massive and that this trend changes for the $\rm 25~M_\odot$ since NH88
find a silicon convective shell ($\rm 1.61~M_\odot$) more external than in our case ($\rm 1.54~M_\odot$) 
in spite of a smaller oxygen exhausted core.
WW95 obtain iron core masses systematically larger than the ones we find (but in the 15 $\rm M_\odot$), 
occurrence which is consistent with the fact that they obtain in general core masses larger than ours.

A comparison between Figure \ref{figmassye} and Figures 2a, 2b, 2c and 2d in \cite{TNH96} (1996) allow also
the checking of the $Y_{\rm e}$ profile just outside the edge of the iron core: also in this case significant
differences are present. In particular:
1) the outer edge of the partially neutronized region in the $\rm 13~M_\odot$ model computed by NH88 is more external
in mass ($\rm \simeq1.5~M_\odot$) and $Y_{\rm e}$ is lower ($Y_{\rm e}\simeq0.491$) than in our corresponding model,
probably because of a more extended O convective shell;
2) a partially neutronized region is preserved between 1.31 and 
$\rm \simeq1.38~M_\odot$ in the $\rm 15~M_\odot$ by NH88, 
at variance with our corresponding model in which, as discussed above, $Y_{\rm e}$ 
jumps from 0.486, within the iron core, to 0.498 outside; this occurrence should be the consequence
of a less extended Si convective which is not able, as in our model, to ingest all the O convective shell;
3) in the $\rm 20~M_\odot$ by NH88 the outer edge of the partially neutronized region is more internal 
($\rm \simeq1.64~M_\odot$) and $Y_{\rm e}$ is lower ($Y_{\rm e}\simeq0.494$) than in our corresponding model;
in this case, at variance with the $\rm 13~M_\odot$, this should be the consequence of a less extended
O convective shell;
4) the $\rm 25~M_\odot$ by NH88 does not show any partially neutronized region, at variance with our corresponding
model; this occurrence happens probably because of an extended
Si convective shell which is able to ingest all the O convective shell, 
similarly to what happens in our $\rm 15~M_\odot$ model.
The impact of these differences on the final explosive nucleosynthesis will be addressed in a forthcoming paper.

A comparison of the presupernova nucleosynthesis with similar data available in literature
is a difficult task because neither NH88 nor WW95 present the preexplosive yields of their models
(indeed WW95 give the preexplosive yields
only in one case, i.e., the $\rm 25~M_\odot$ of solar chemical composition).
A comparison with ABP96 is not meaningful
because they stop their computation at the central Ne exhaustion. Note that they claim that 
above $\rm 1.9~M_\odot$, in the $\rm 20~M_\odot$ model, the chemical composition does not change
significantly because of the very short evolutionary timescales of the more advanced phases,
and hence they compare the abundances of isotopes up to $\rm ^{27}Al$ with WW95 and NH88; actually
we find that isotopes heavier than $\rm ^{17}O$ suffer substantial modifications even
after central Ne exhaustion. 
Table \ref{tabcenam} shows a comparison of the
presupernova yields (in $\rm M_\odot$) of the isotopes lighter than A=27 to the ones computed
at the central Ne exhaustion. In the last column of Table \ref{tabcenam} we report the site in
which the various isotopes are modified after the central Ne exhaustion;
the labels "p" and "d" indicating respectively whether each isotope is produced or destroyed.
\begin{table}
\dummytable\label{tabcenam}
\end{table}
\placetable{tabcenam}

\section{Summary and Conclusions}
In this paper, which is the second of the series, we have presented and discussed 
in some detail the presupernova evolution of four stellar masses having solar chemical
composition. All the various burning phases have been studied by adopting a very extended
network fully embedded into the stellar evolutionary code (FRANEC).
A comparison of our results with similar ones available in the literature shows
that, in spite of an overall similarity, the existing differences are significant and at 
present not completely understood.

\acknowledgements
One of us (A.C.) thanks the Astronomical Observatory of Rome and its Director, Prof. Roberto Buonanno,
for the generous hospitality at Monteporzio Catone. This paper has been partially supported by the 
MURST (COFIN98).

\appendix

\section{Explosive nucleosynthetic yields}
In this appendix we anticipate the elemental explosive nucleosynthetic yields coming from
a set of models including the present ones and those of lower metallicity, i.e.,
$Z=0$ ($Y=0.23$) and $Z=10^{-3}$ ($Y=0.23$). The explosion has been obtained by
depositing at the border of the iron core an amount of energy equal to $\rm 1.2\cdot10^{51}~erg$
plus the binding energy. Each table refers to a single mass and each column refers to
a different choice of the mass cut (i.e., for a different choice of the ejected $\rm ^{56}Ni$).
The last column refers to a mass cut equal to the iron core mass (which also corresponds
to the maximum amount of $\rm ^{56}Ni$ available).
All the unstable isotopes have been decayed into their parent stable nuclei.
These results will be discussed in detail in a forthcoming paper.
\begin{table}
\dummytable\label{A1}
\end{table}
\placetable{A1}
\begin{table}
\dummytable\label{A2}
\end{table}
\placetable{A2}
\begin{table}
\dummytable\label{A3}
\end{table}
\placetable{A3}
\begin{table}
\dummytable\label{A4}
\end{table}
\placetable{A4}
\begin{table}
\dummytable\label{A5}
\end{table}
\placetable{A5}
\begin{table}
\dummytable\label{A6}
\end{table}
\placetable{A6}
\begin{table}
\dummytable\label{A7}
\end{table}
\placetable{A7}
\begin{table}
\dummytable\label{A8}
\end{table}
\placetable{A8}
\begin{table}
\dummytable\label{A9}
\end{table}
\placetable{A9}
\begin{table}
\dummytable\label{A10}
\end{table}
\placetable{A10}
\begin{table}
\dummytable\label{A11}
\end{table}
\placetable{A11}

\newpage

\centerline{\bf FIGURE CAPTIONS}
\vspace{1cm}

\figcaption[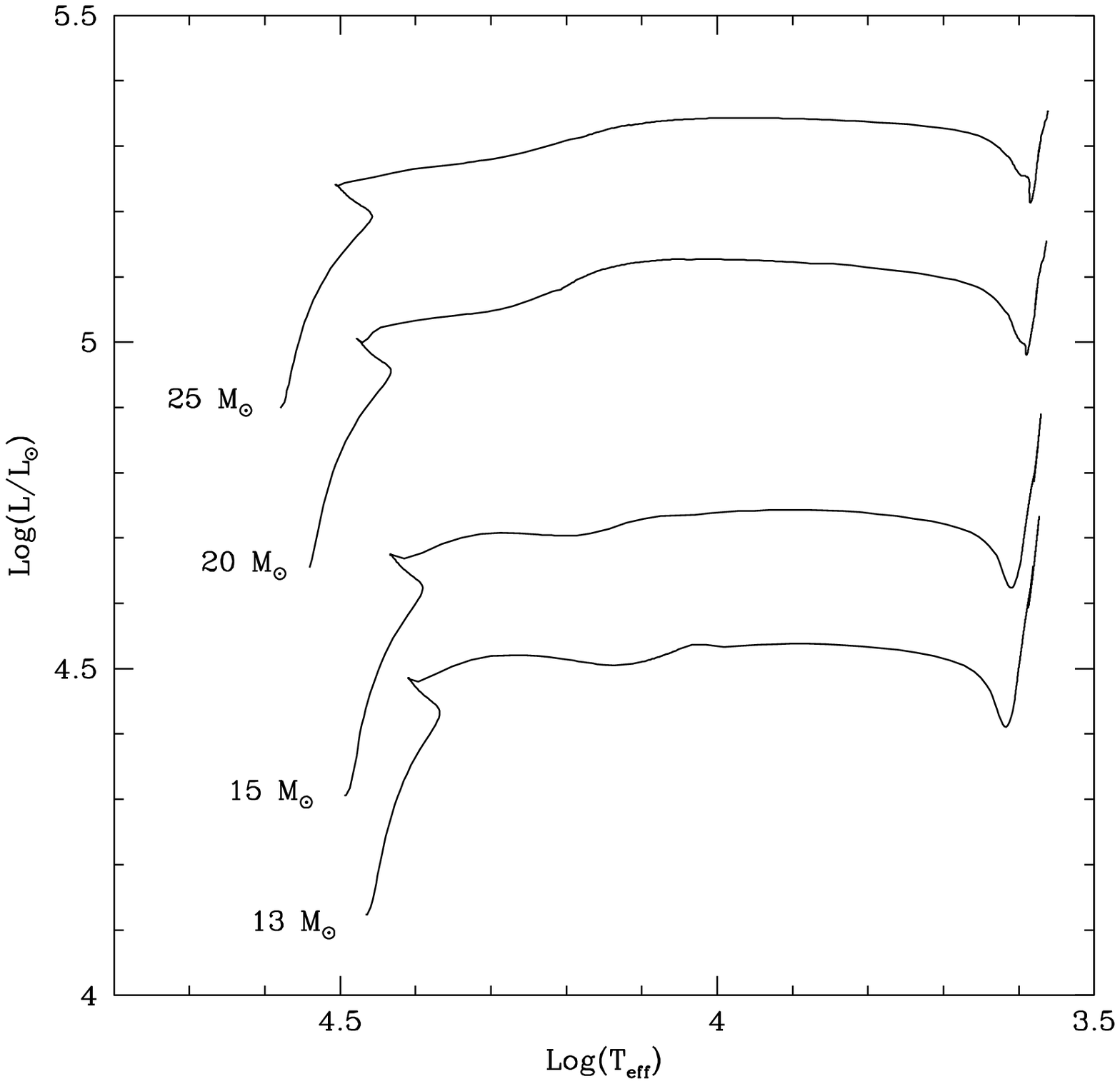]{Path followed by the computed models in the HR diagram.\label{fighr}}

\figcaption[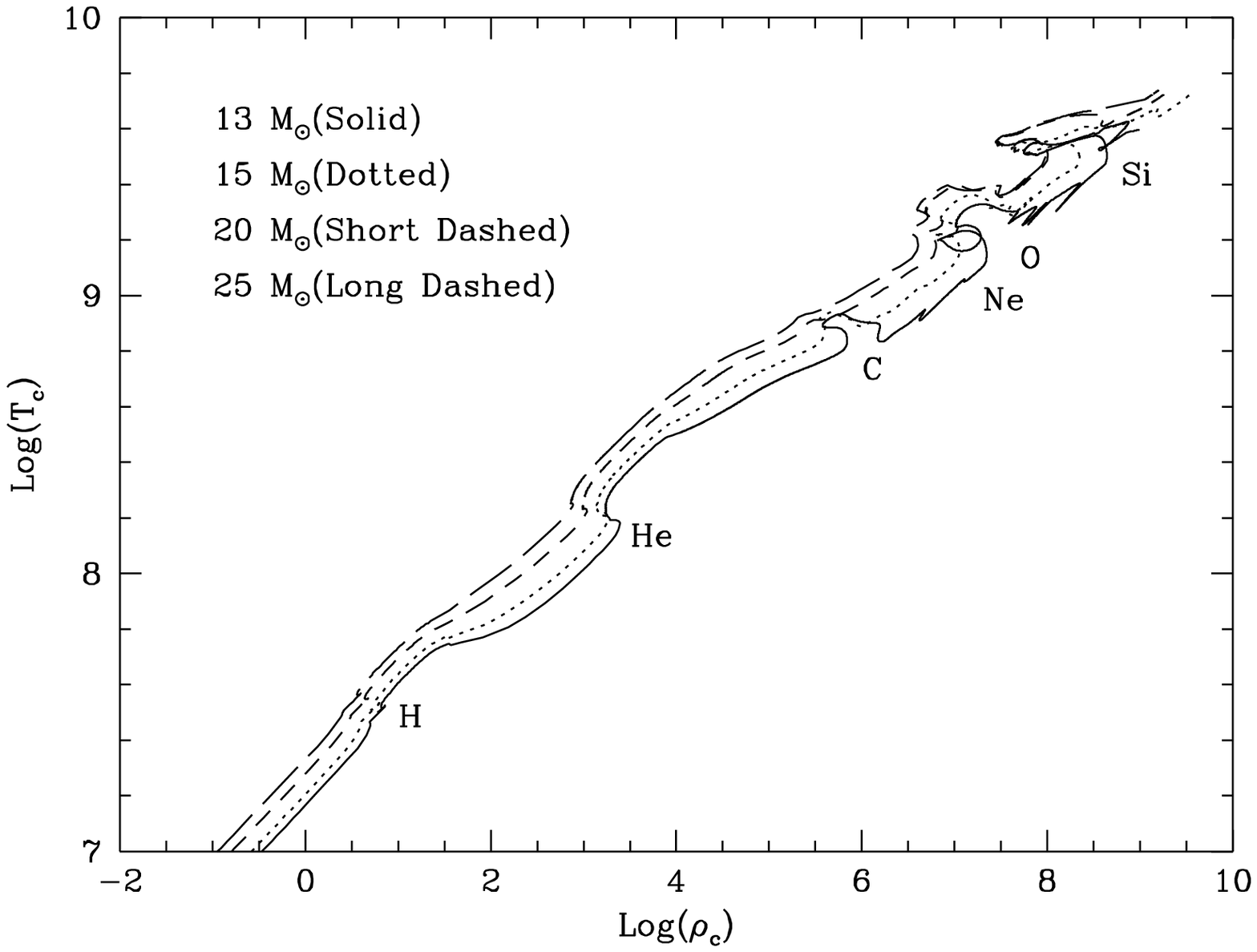]{Path followed by the computed models in the ${\rm Log}T_{\rm c}-{\rm Log}\rho_{\rm c}$ diagram;
the various nuclear burning ignitions are shown in the figure. \label{figtcroc}}

\figcaption[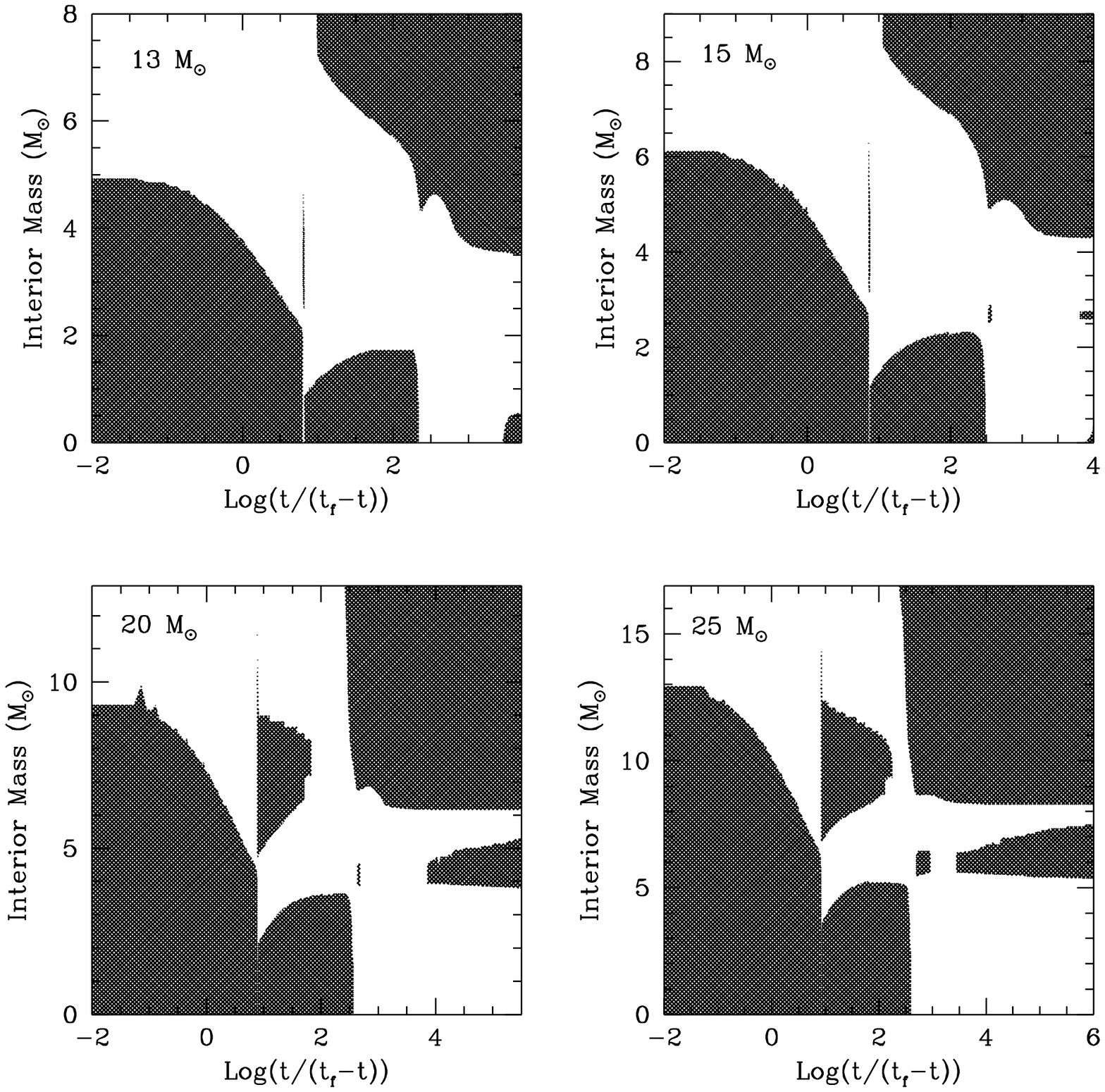]{Temporal behaviour of the convective zones for the computed models from the MS phase up to the central 
carbon ignition.\label{figconv1}}

\figcaption[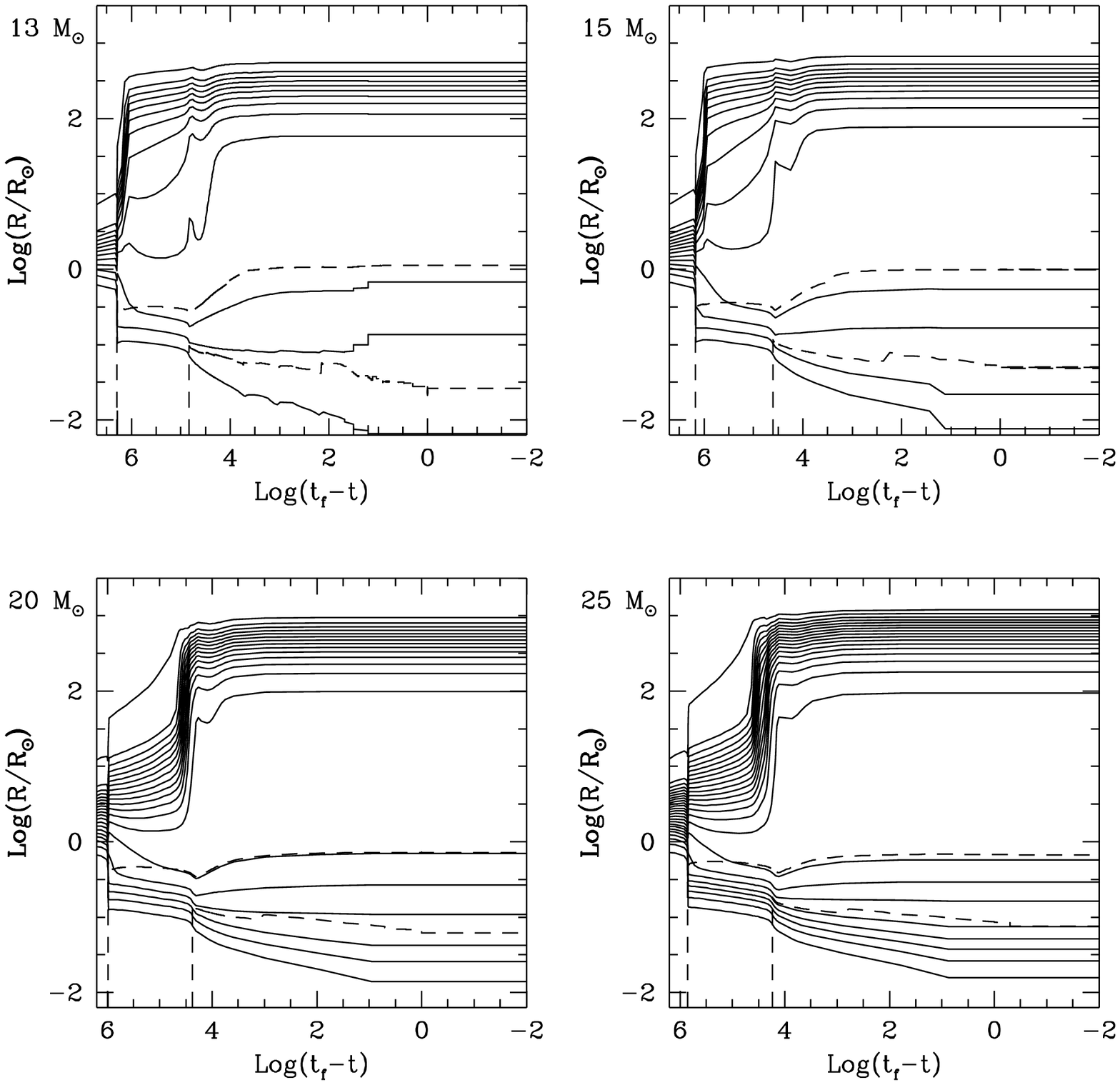]{Temporal evolution of the radius of the internal mass layers in steps of
$\rm 1~M_\odot$ from the H exhaustion to the presupernova stage for the four computed models. The dashed
lines refer to the mass location of the H (the more external one) and of the He (the more internal one) burning 
shells. \label{figragint}}

\figcaption[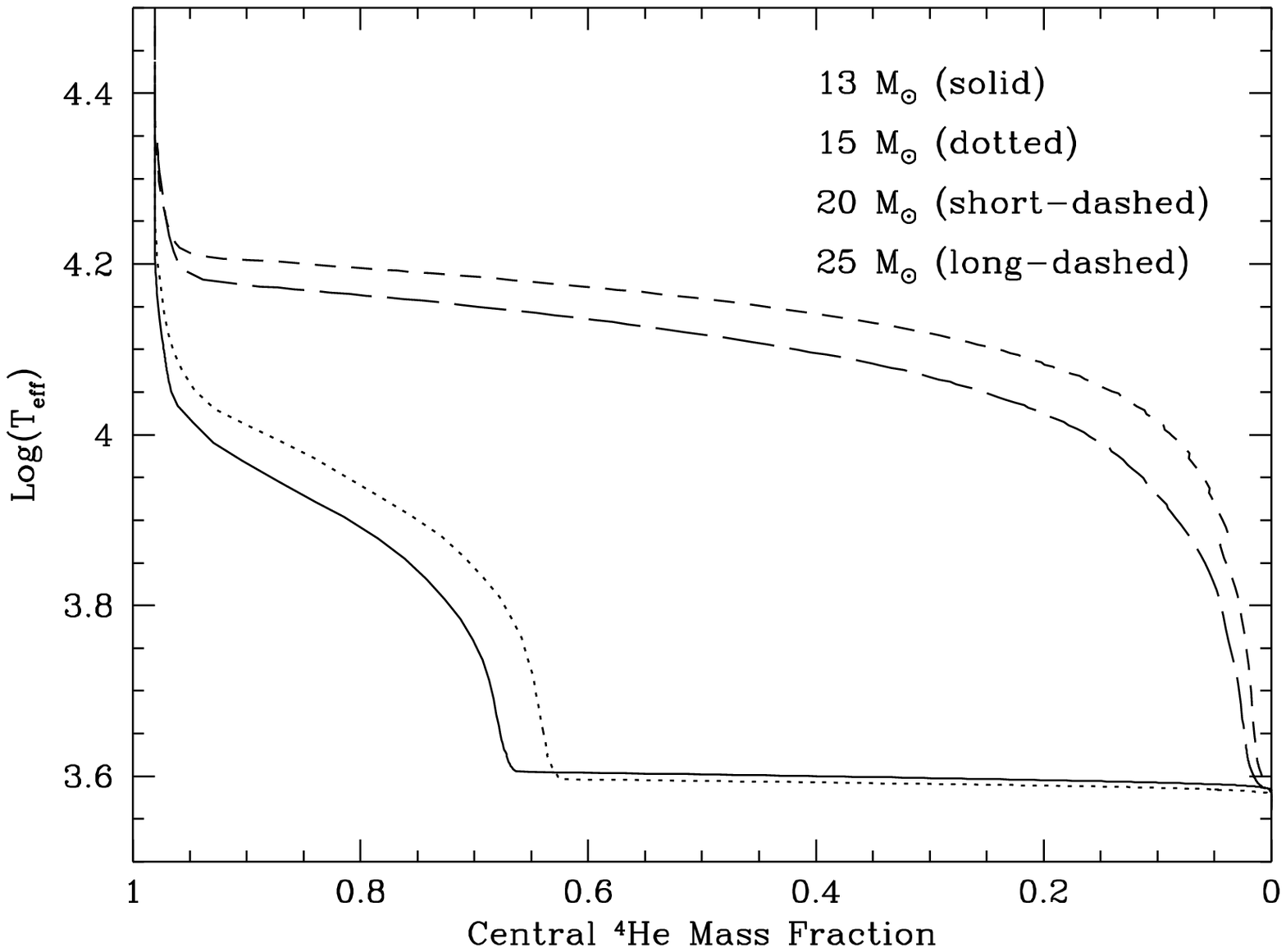]{Temporal evolution of the surface effective temperature as a function of the central He mass
fraction.\label{fighecteff}}

\figcaption[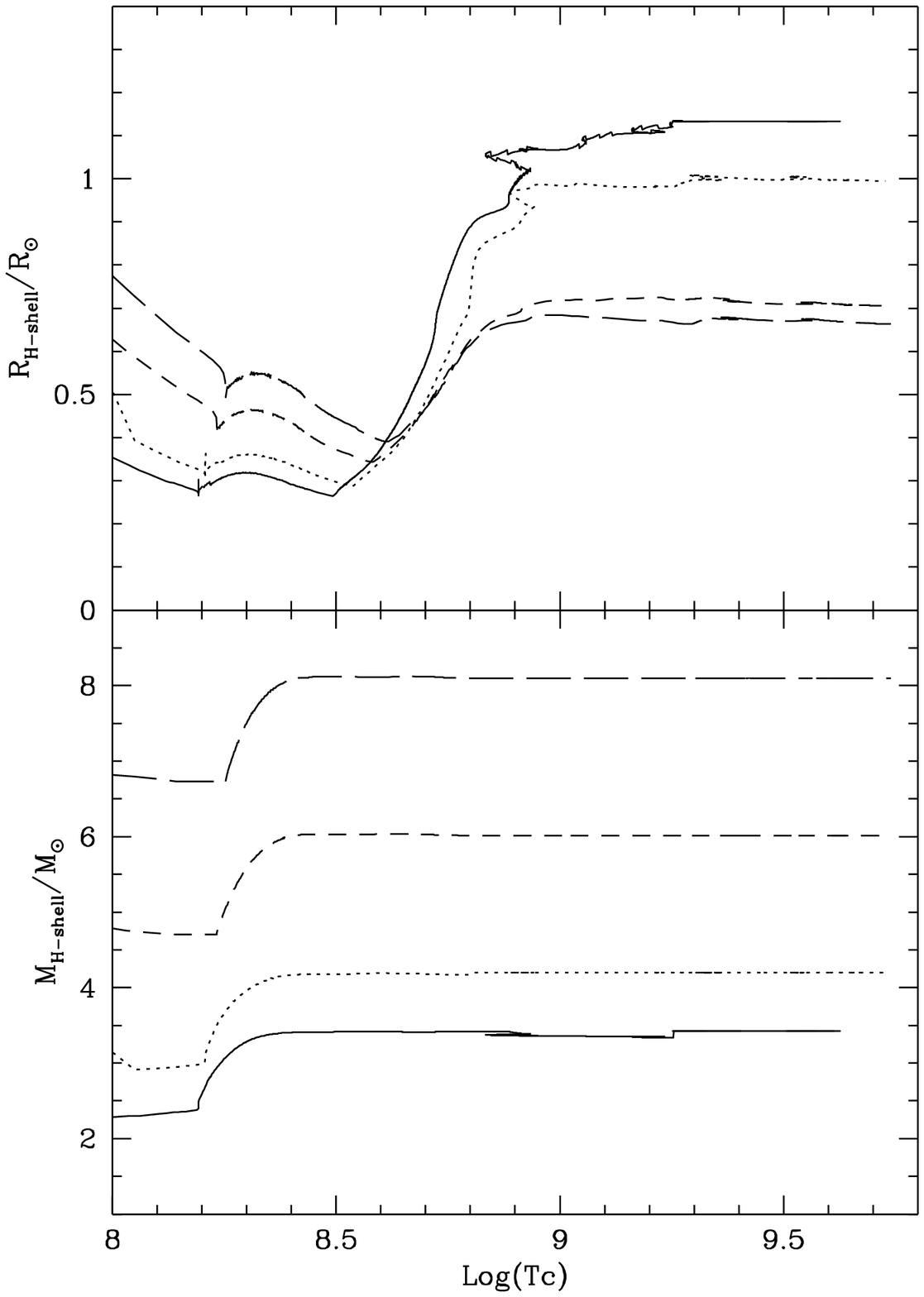]{Evolution of both the radius ({\em upper panel}) and the mass ({\em lower panel})
of the H burning shell (defined as the mass corresponding to the maximum energy generation rate
provided by the H burning reactions) as a function of the central temperature up to the
presupernova stage. The {\em solid lines} refer to the $\rm 13~M_\odot$ model, the {\em dotted lines}
to the $\rm 15~M_\odot$ model, the {\em dashed lines} to the $\rm 20~M_\odot$ model and
the {\em long dashed lines} to the $\rm 25~M_\odot$ model.\label{figmrshell}}

\figcaption[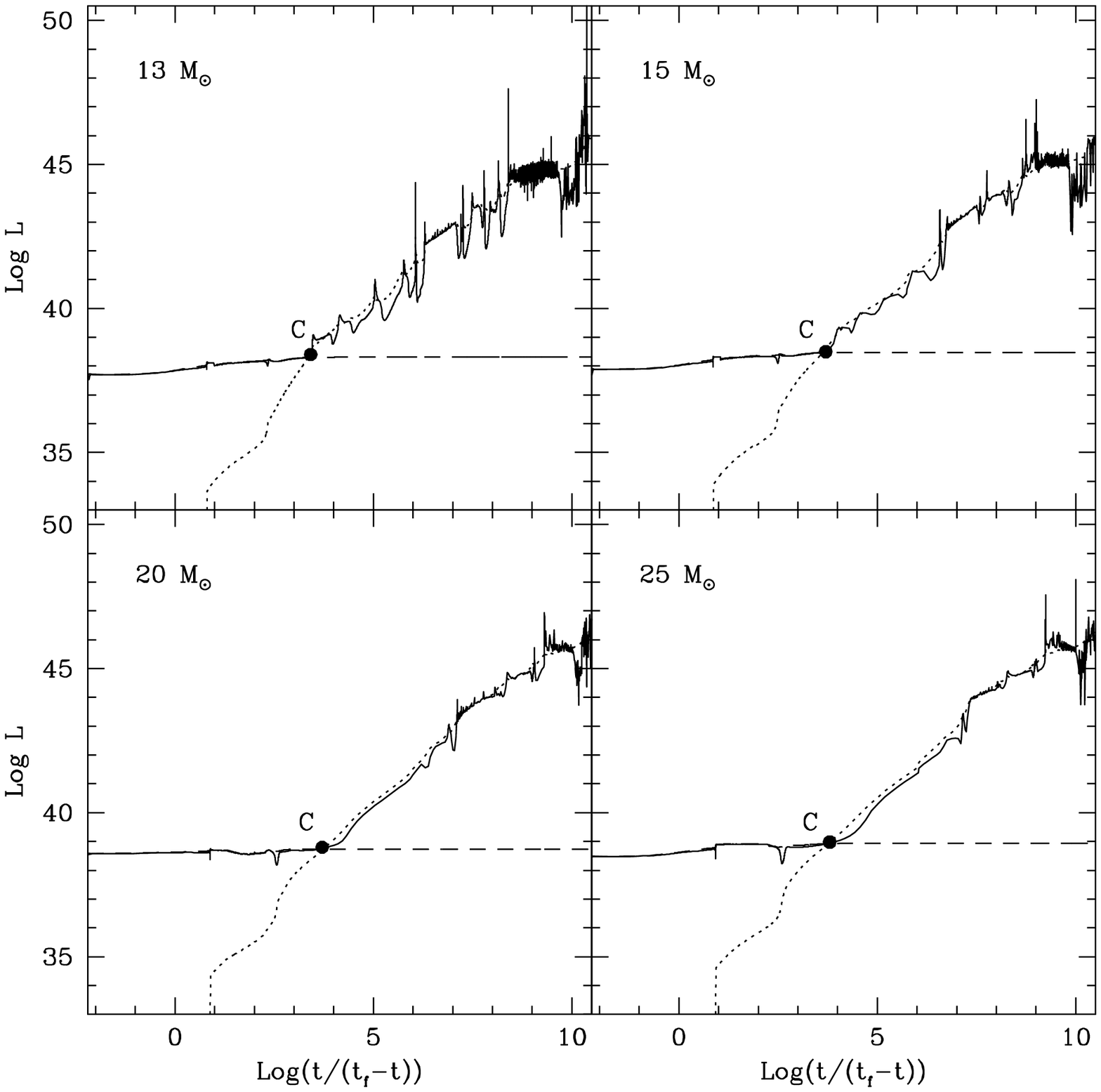]{Temporal behaviour of the various luminosities: photons ({\em dashed lines}), neutrinos
({\em dotted lines}) and nuclear ({\em solid lines}). The carbon burning ignition is indicated in each
panel by the big filled dot labeled as "C". \label{figvarielum}}

\clearpage

\figcaption[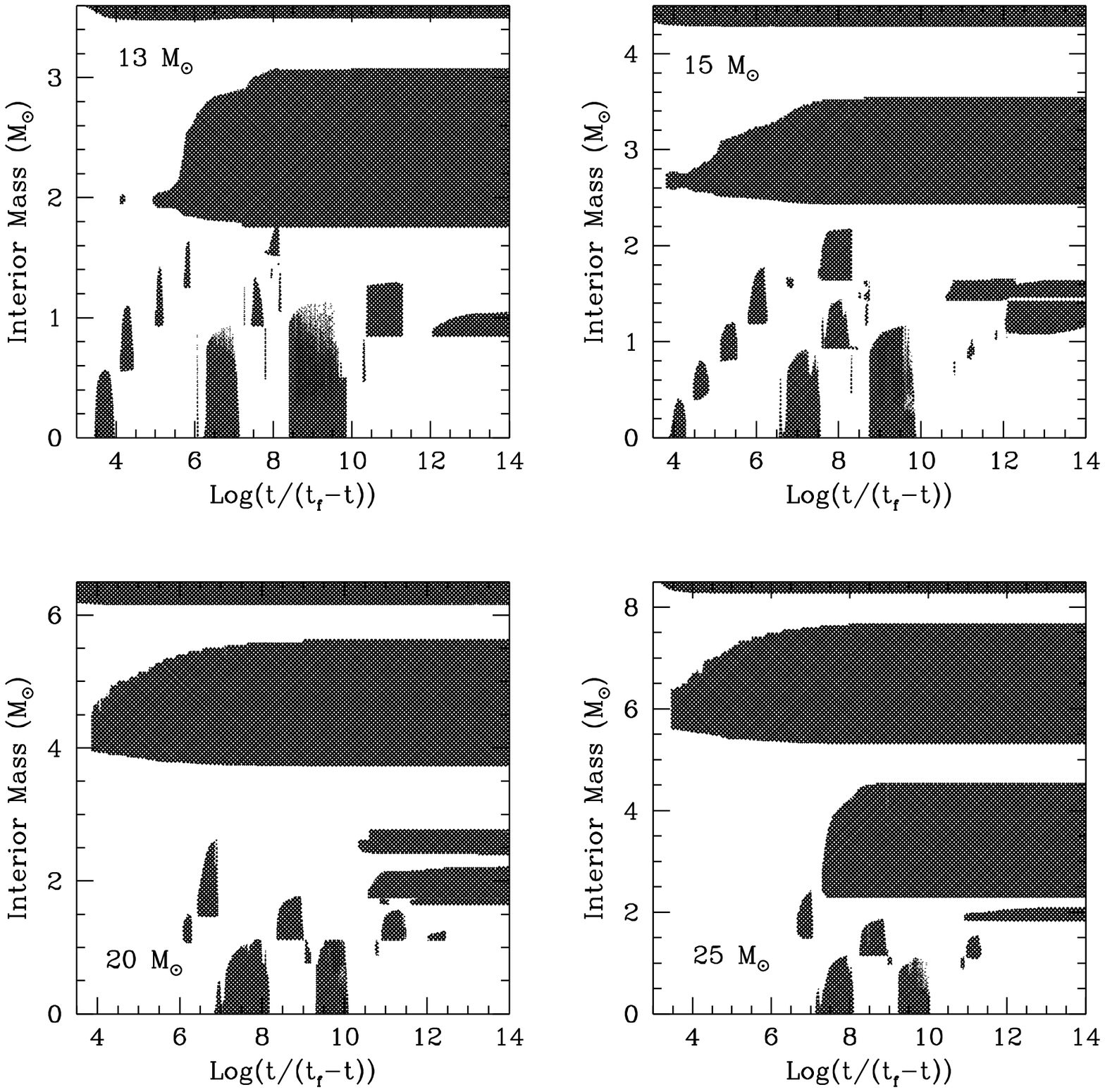]{Temporal behaviour of the convective zones for all the computed models from the central He
exhaustion up to the presupernova stage.\label{figconv2}}

\figcaption[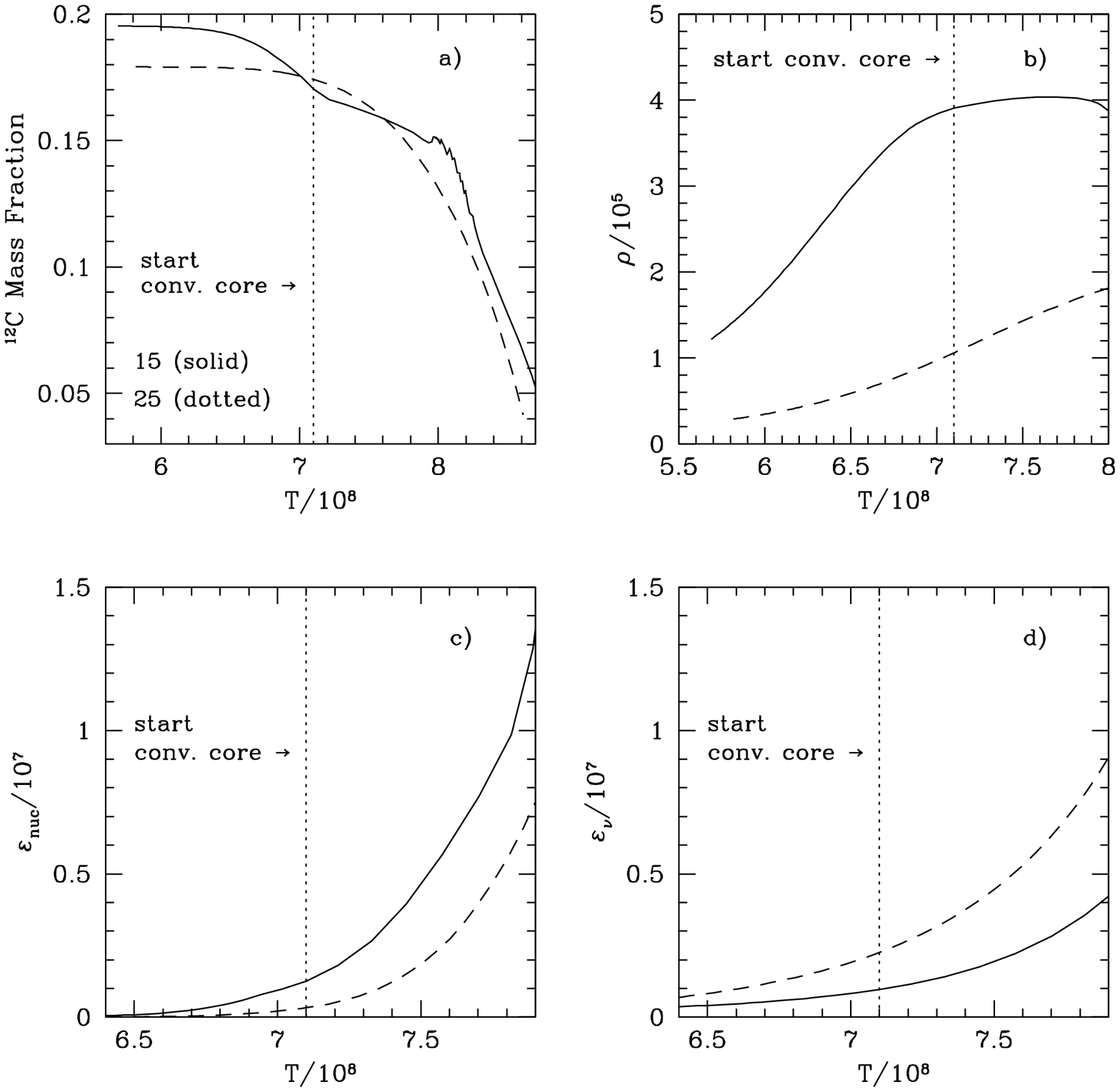]{
Evolution of central $\rm ^{12}C$ mass fraction ({\em panel a}), central density ({\em panel b}),
central nuclear generation rate ({\em panel c}) and neutrino losses ({\em panel d})
as a function of the central temperature
for both the 15 ({\em solid line}) and the 25 $\rm M_\odot$ ({\em dashed line})
during central carbon burning. The vertical {\em dotted line} marks the 
formation of the convective core in the $\rm 15~M_\odot$.\label{figcburnprop}}

\figcaption[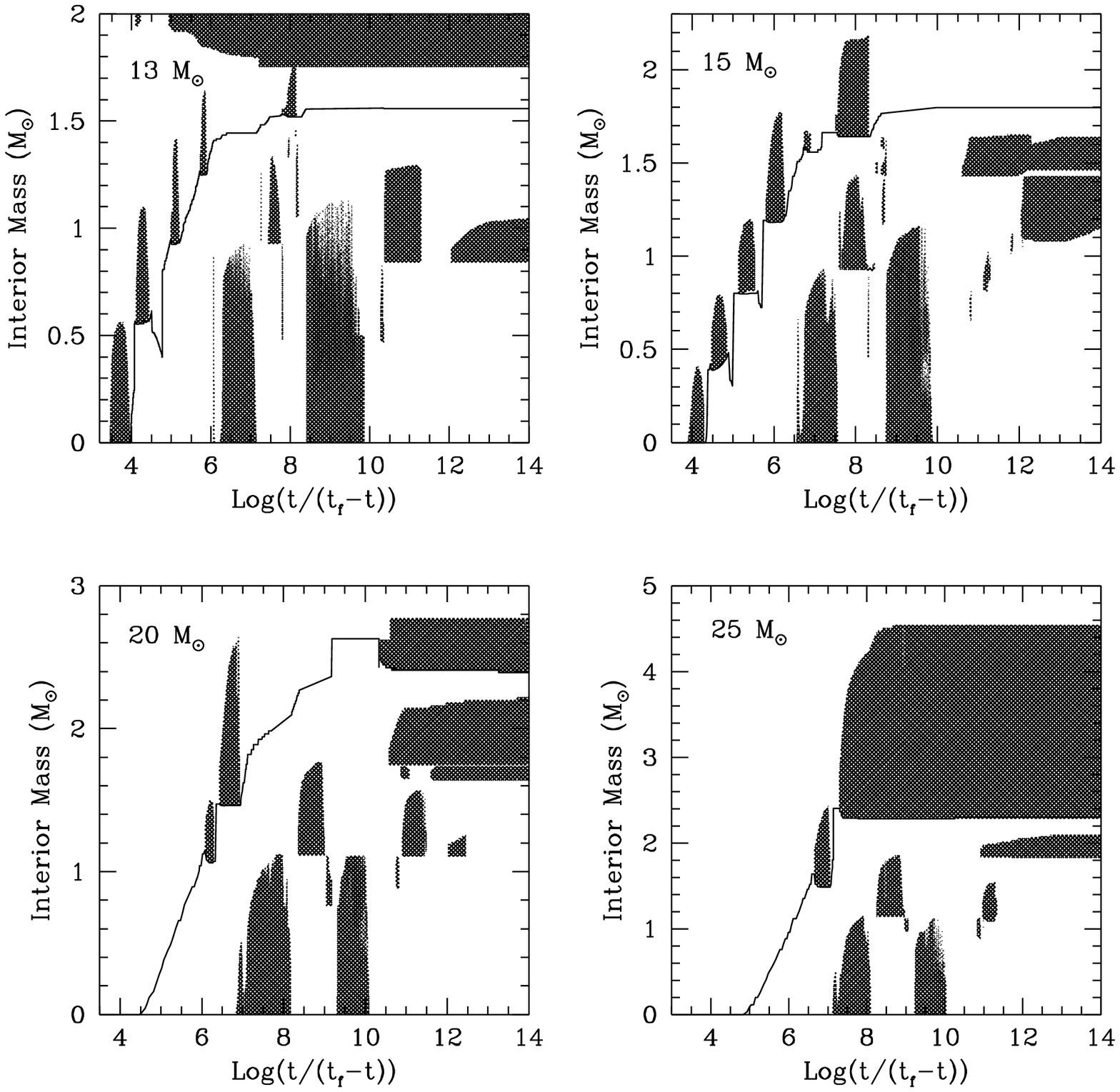]{Evolution of the mass location of the C burning shell superimposed to the
convective zones as a function of time from central C burning to the presupernova stage.
\label{figcshellprop}}

\figcaption[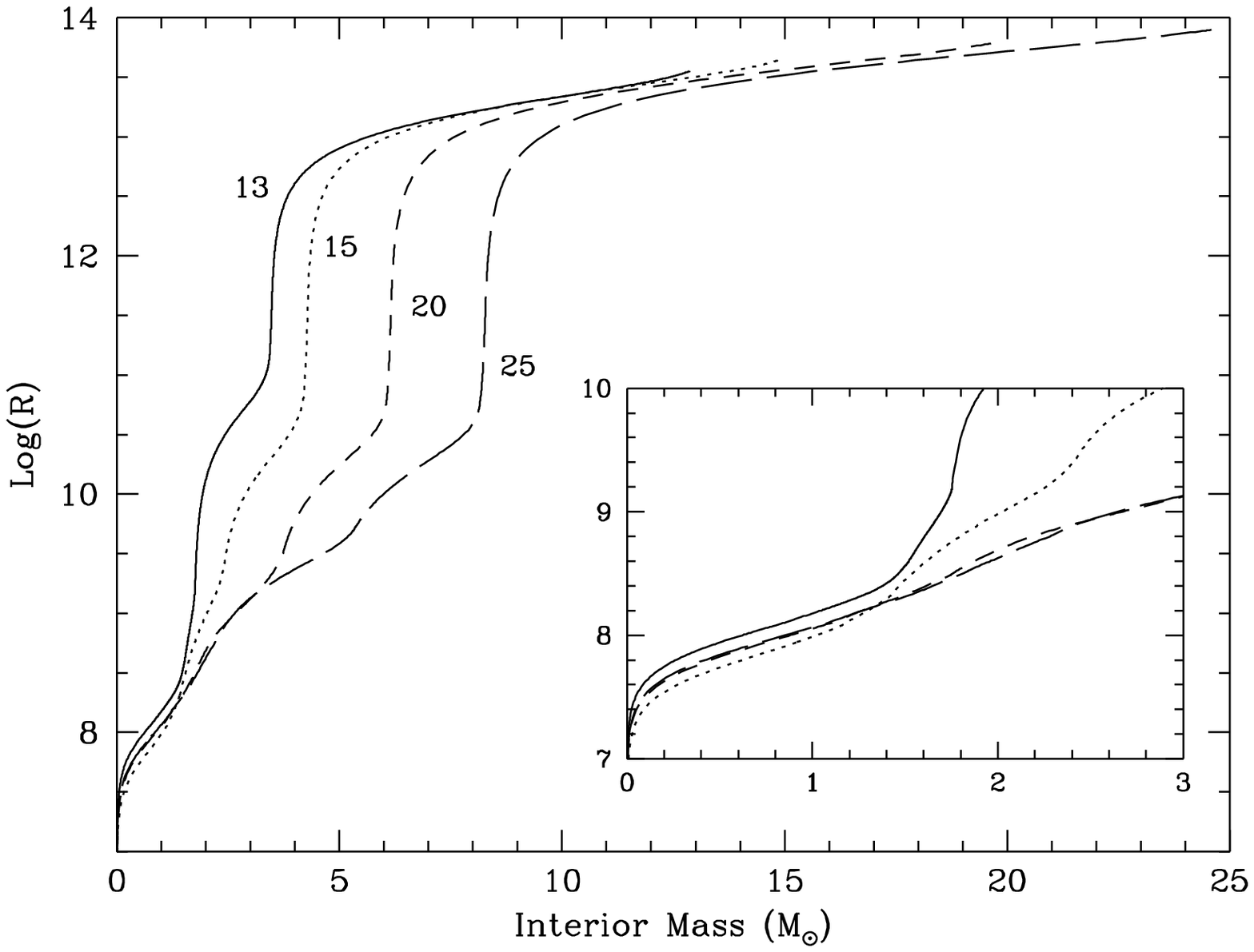]{Presupernova radial profile as a function of the interior mass for all
the computed models.\label{figmassrag}}

\figcaption[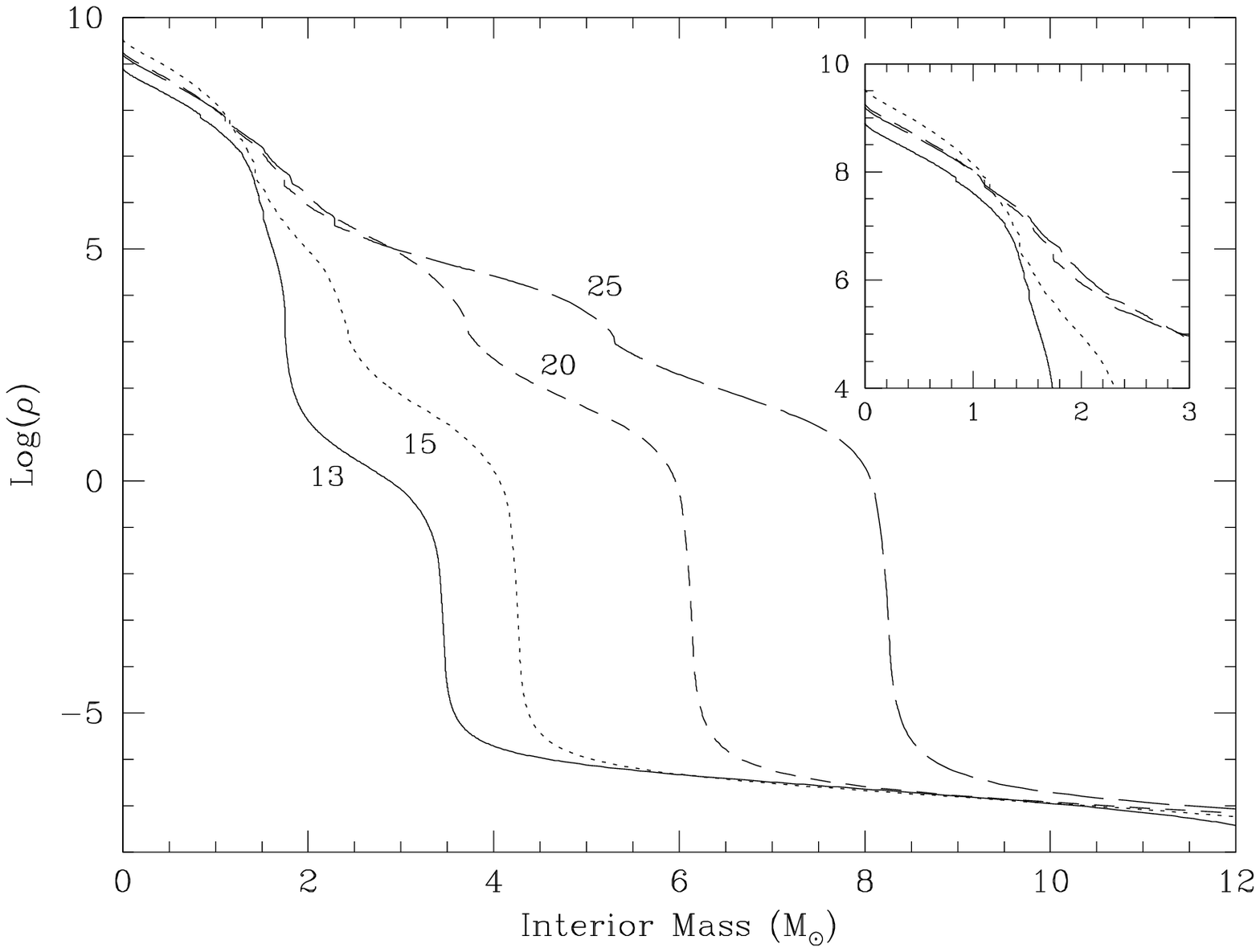]{Presupernova density profile as a function of the interior mass for all
the computed models.\label{figmassdens}}

\figcaption[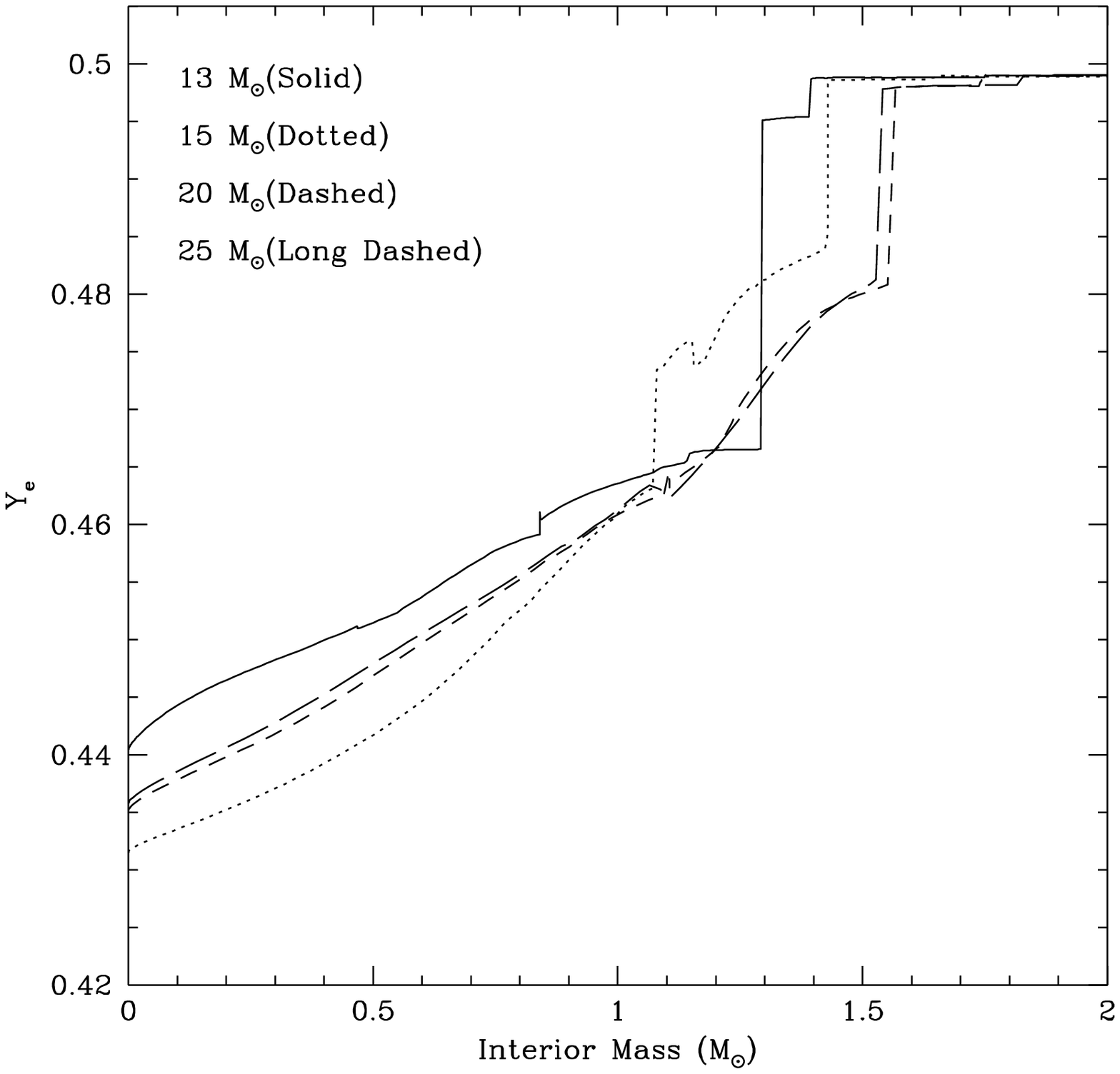]{Presupernova electron fraction profile as a function of the interior mass
within the inner $\rm 2~M_\odot$ for all the computed models.\label{figmassye}}

\figcaption[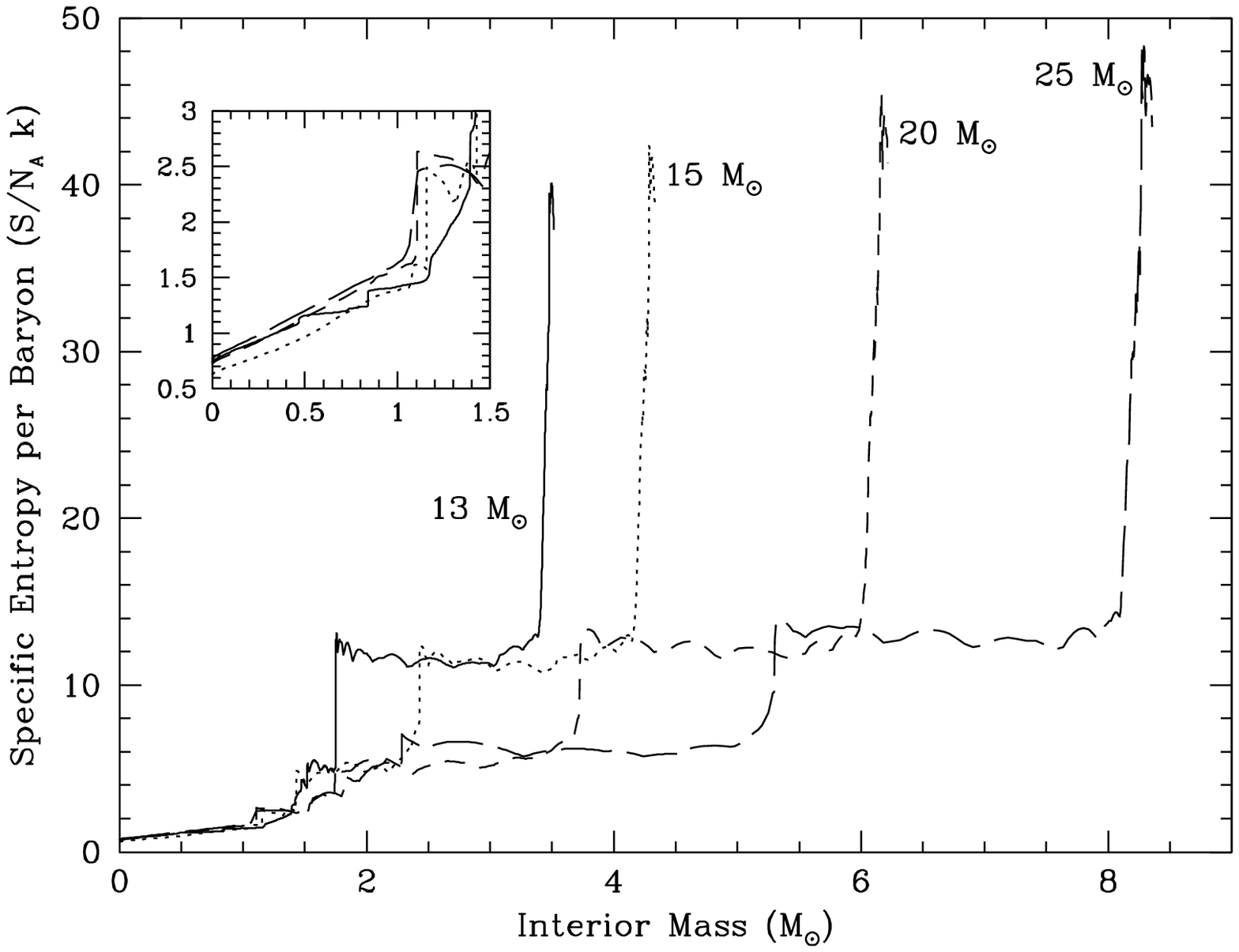]{Presupernova profile of the specific entropy per barion as a function of the
interior mass. Note that the specific entropy per barion is computed only for temperatures
larger than $\rm 10^{6}~K$, where the matter is assumed to be completely ionized.
\label{figmassentr}}

\clearpage

\figcaption[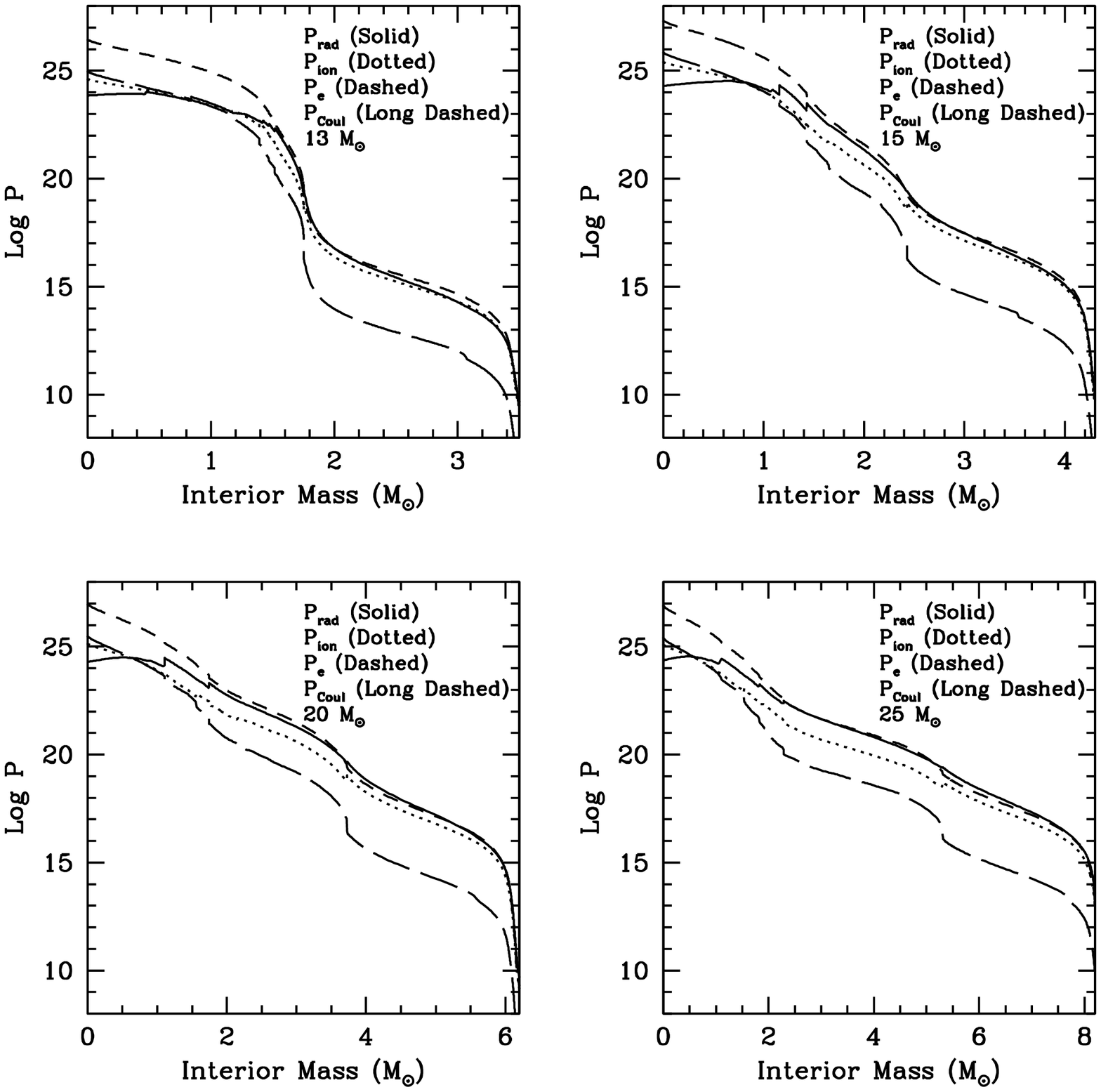]{Various contributions to the total pressure within the
He core of all the presupernova models: radiation ({\em solid lines}),
ions ({\em dotted lines}), electrons ({\em dashed lines}), Coulomb
corrections ({\em long dashed lines}).\label{figvariepress}}

\figcaption[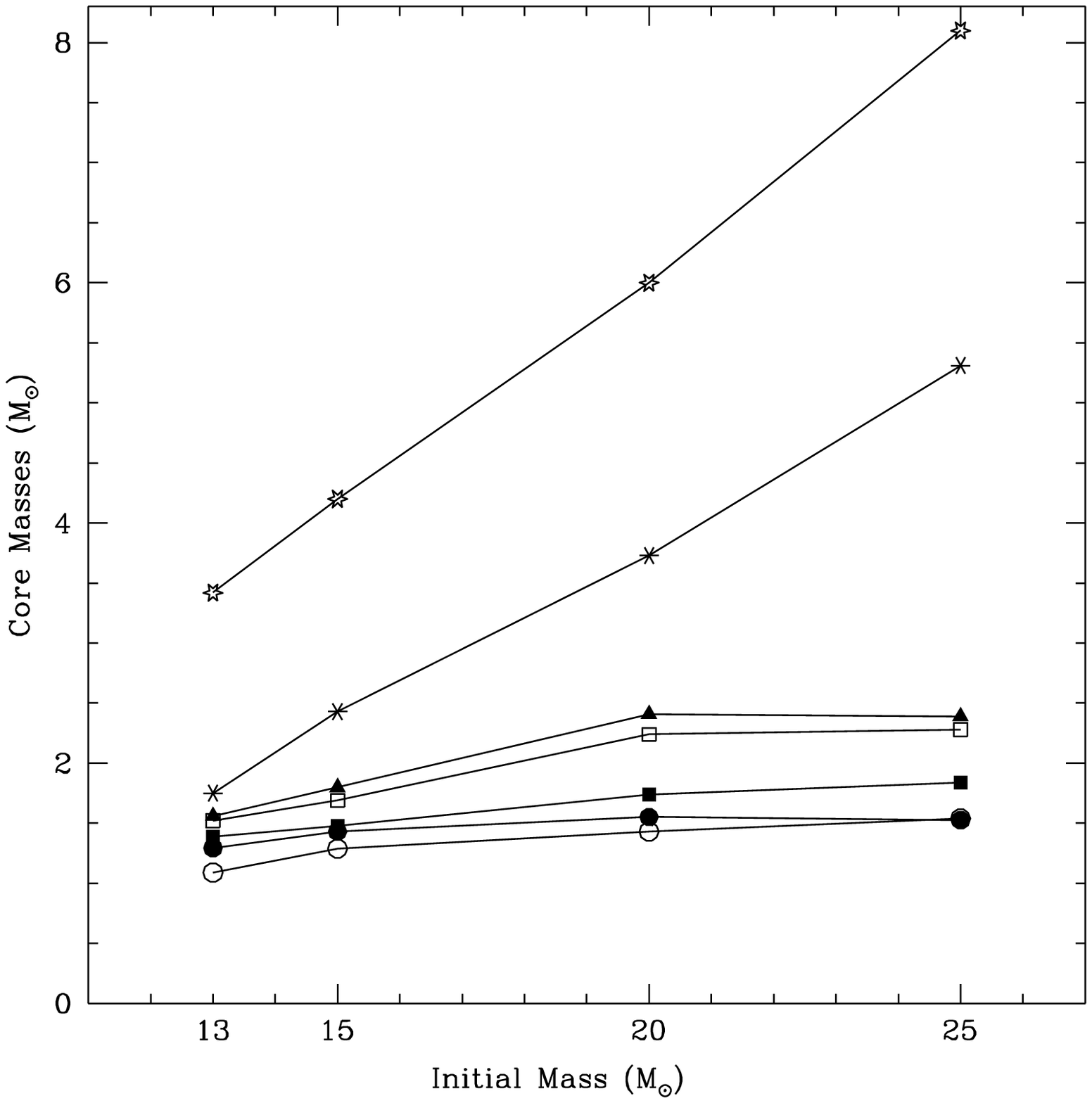]{Final location of the various core masses (or equivalently of the
various nuclear burning shells) for all the computed models: 
H ({\em open stars}), He ({\em skeletal stars}), C ({\em filled triangles}),
Ne ({\em open squares}), O ({\em filled squares}), Si ({\em filled circles}),
Iron Cores ({\em open circles}).
\label{figmcores}}

\figcaption[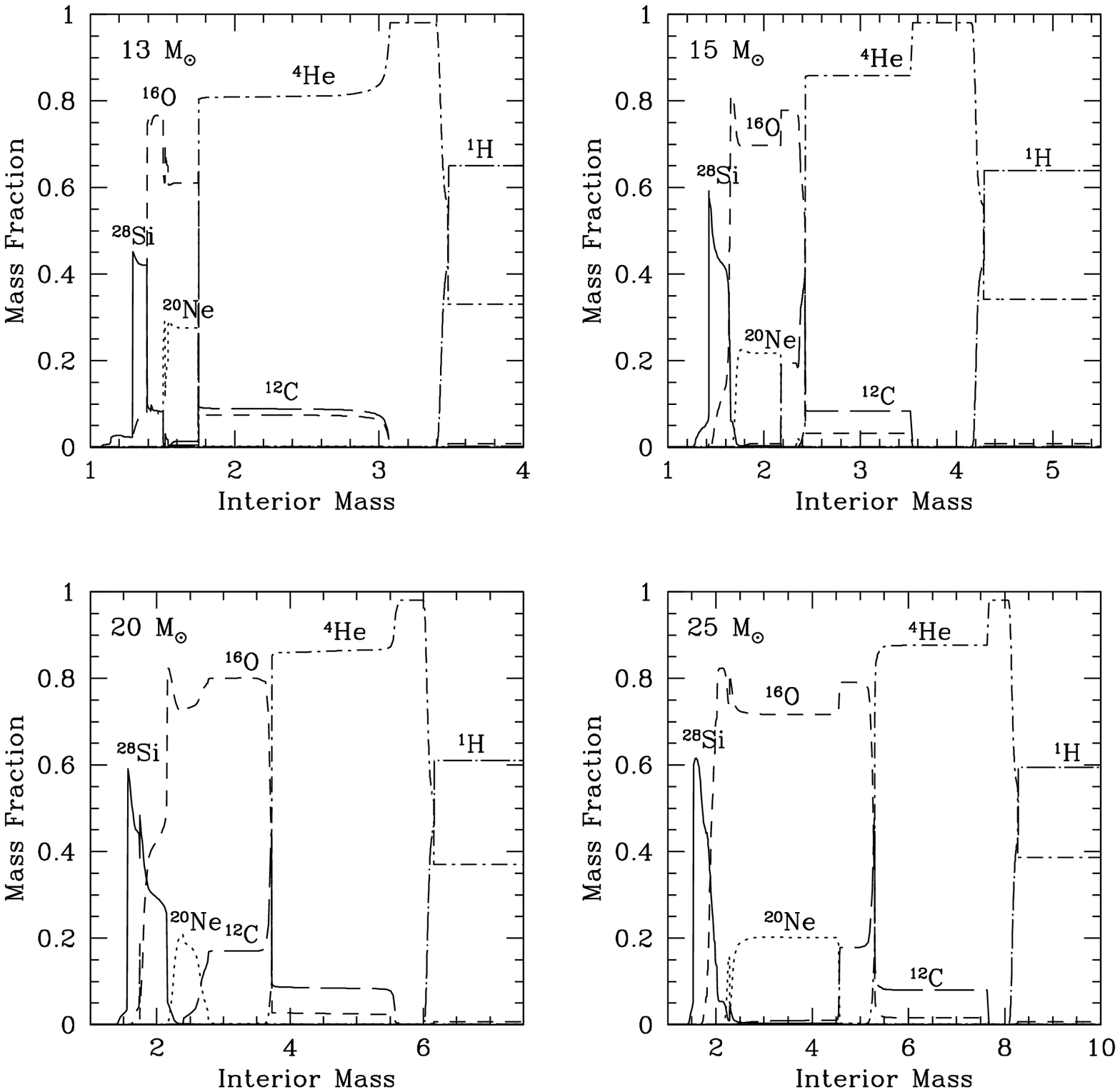]{Internal profiles of the most abundant nuclear species for all the presupernova models.
\label{figmajors}}

\figcaption[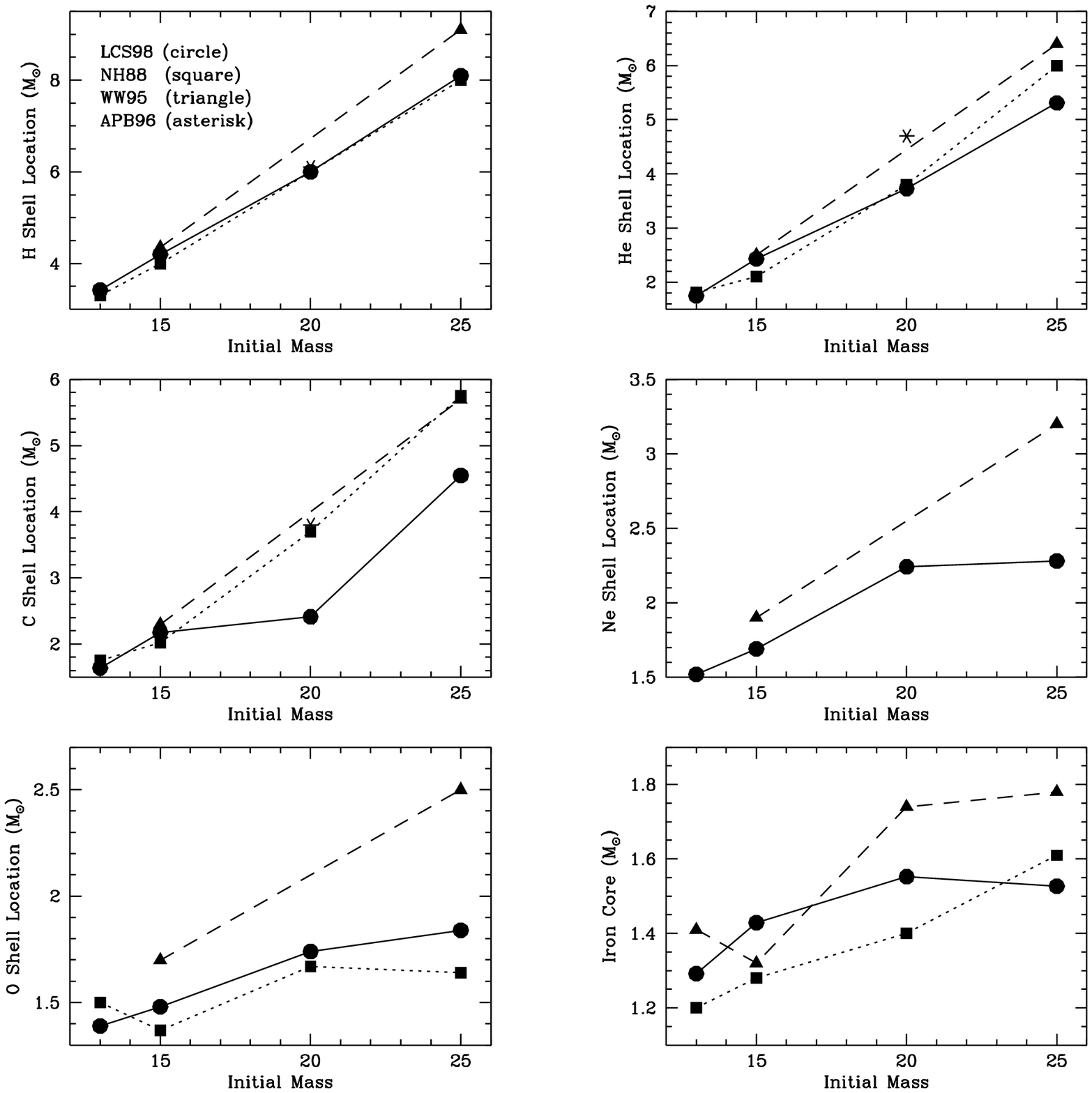]{Comparison of the final location in mass of the various nuclear burning shells of all
the presupernova models with WW95 (filled triangles), NH88 (filled squares) and APB96 (asterisk).
\label{figcompmcore}}

\end{document}